\documentclass{aa}  

\usepackage{graphicx}
\usepackage{txfonts}
\usepackage{subcaption}       

\newcommand{\msun}{$M_{\odot}$}
\newcommand{\rsun}{$R_{\odot}$}
\newcommand{\lsun}{$L_{\odot}$}

\newcommand{\teff}{$T_\mathrm{eff}$}
\newcommand{\logg}{$\log g$}
\newcommand{\kms}{km\,s$^{-1}$}
\newcommand{\BD}{BD+20~5391}

\newcommand{\highlight}[1]{{#1}}

\begin{document}

   \title{The twin red giant branch system BD+20 5391}

   \subtitle{A case study of low-mass double-core evolution}

   \titlerunning{The twin RGB system BD+20 5391. }

   \authorrunning{M. Kurpas, M. Dorsch, S. Geier, B. Kubátová et al.}

   \author{M. Kurpas\inst{1},
   M. Dorsch\inst{1},
   S. Geier\inst{1},
   B. Kubátová\inst{2},
   J. Vos\inst{2},
   M. Cabezas\inst{2},
   E. Kundra\inst{3},
   J. Budaj\inst{3},
   K. Deshmukh\inst{4,5},
   V. Schaffenroth\inst{6},
   I. Pelisoli\inst{7},
   H. Dawson\inst{1},
   M. Pritzkuleit\inst{1},
   O. Maryeva\inst{2},
   J. Kub\'at\inst{2}
        }

   \institute{Institut für Physik und Astronomie, Universität Potsdam, Haus 28, Karl-Liebknecht-Str.\ 24/25, 14476 Potsdam, Germany\\
             \email{marie.scheffen@uni-potsdam.de}
            \and Astronomical Institute AS CR, Fričova 298, 251 65 Ondřejov, Czech Republic
            \and Astronomical Institute of the Slovak Academy of Sciences, 059 60 Tatranská Lomnica, Slovakia
            \and Institute of Astronomy, KU Leuven, Celestijnenlaan 200D, 3001 Leuven, Belgium
            \and Leuven Gravity Institute, KU Leuven, Celestijnenlaan 200D, box 2415, 3001 Leuven, Belgium
            \and Thüringer Landessternwarte Tautenburg, Sternwarte 5, 07778 Tautenburg, Germany
            \and University of Warwick, Department of Physics, Gibbet Hill Road, Coventry, CV4 7AL, UK\\ }

   \date{Accepted August 10, 2025}

\abstract
  {Understanding interactions of binary systems on the red giant branch is crucial to understanding the formation of compact stellar remnants such as helium-core white dwarfs (He-WDs) and hot subdwarfs. However, the detailed evolution of such systems, particularly those with nearly identical components, remains under-explored.} 
  {We aim to analyse the double-lined spectroscopic binary system BD+20 5391, composed of two red giant stars, in order to characterise its orbital and stellar parameters and to constrain its evolution.} 
  { Spectroscopic data were collected between 2020 and 2025 using the Ond\v{r}ejov Echelle Spectrograph and the Mercator Échelle Spectrograph. The time-resolved spectra were fitted with models to determine the radial velocity curve and derive the system's parameters. We then used the position of both stars in the Hertzsprung-Russell diagram to constrain the system's current evolutionary state, and we discuss potential outcomes of future interactions between the binary components.} 
  {We find that the two stars in BD+20 5391 will likely initiate Roche lobe overflow (RLOF) simultaneously, leading to a double-core evolution scenario. The stars’ helium core masses at RLOF onset will be almost identical, at 0.33\,$\mathrm{M}_{\odot}$. This synchronised evolution suggests two possible outcomes: common envelope ejection, resulting in a short-period double He-WD binary, or a merger without envelope ejection. In the former case, the resulting double He-WD may merge later and form a hot subdwarf star. } 
  {This study provides a valuable benchmark example for understanding the evolution of interacting red giant binaries, which will be discovered in substantial numbers in upcoming large-scale spectroscopic surveys. 
  }


   \maketitle

\section{Introduction}\label{intro}

Understanding binary stellar evolution is fundamental to our understanding of stellar populations, given that a significant fraction of stars form in binaries or higher-order systems. Of the known F-, G-, and K-type stars, more than 40\% reside in multiple-star systems \citep{Duchene2013,Offner2023}.
However, many aspects of binary interaction physics remain poorly understood, even though they can have a profound impact on stellar evolution. 

Some stellar classes, such as Be stars \citep{Be}, helium-core white dwarfs \citep[He-WDs; e.g.][]{Li2019}, R Coronae Borealis stars \citep{RCB2012}, magnetic Ap stars \citep{Ferrario2009,Tutukov2010}, and hot subdwarfs \citep[sdO/B,][]{Heber2024}, are thought to form exclusively through binary evolution channels. Establishing connections between progenitor systems and their post-interaction products is essential for advancing our understanding of binary interactions and, by extension, stellar evolution.

In low-mass binary systems, significant interaction often begins once one or both components ascend the red giant branch~(RGB). If a red giant fills its Roche lobe in a sufficiently wide orbit, stable Roche lobe overflow (RLOF) can strip it to a helium core, yielding a He-WD in a wide binary system \citep[e.g.][]{Iben1984,Nelemans2005}, or a hot subdwarf star if helium is ignited. 
At high mass ratios, the mass transfer is unstable and a common envelope (CE) is formed, which can be ejected and leave a He-WD or sdB remnant in a short-period binary \citep{Han2002}. 

In the extreme case that the two stars in a binary system simultaneously evolve into red giants, double-core CE evolution  can occur \citep[e.g.][]{Paczynski1976, Brown1995, Belczy_ski2001, Dewi2006, Justham_HerichsdO}. In this scenario, the expanding envelopes of the two stars form a CE as their cores spiral inwards. 
The parameter space that allows for double-core CE evolution is highly constrained. 
The two stars must have nearly equal initial masses so that the secondary has already evolved off the main sequence by the time the primary initiates RLOF, which must happen before it reaches the RGB tip \citep{Dewi2006}. 
In this case, stable mass transfer initially occurs; however, as the secondary expands in response to accretion, it ultimately overfills its Roche lobe and the two stars enter a CE phase simultaneously. 

If the envelope is successfully ejected, the result is a compact binary composed of the exposed cores of both stars. Although the double-core CE scenario has not been directly confirmed observationally, compact double-core binaries like the double hot subdwarf system PG~1544+488 \citep{PG15disc, Sener2014} are known.

Binaries containing two red giant stars of similar mass (`twin red giants') are exceedingly rare. Previous research on such objects was focused on oscillating RGB stars \citep{doubleRGosc, Beck2018, Beck2022, Beck2024, Brogaard2022}, but only two -- KIC~9163796 \citep{Beck2018} and KIC~4054905 \citep{Brogaard2022} -- feature components with nearly identical masses.
\highlight{\citet{Uzundag2022} compiled a volume-limited sample of low-mass red giants within 500 pc of the Sun that meet the criteria for potential hot subdwarf progenitors, using data from the second \textit{Gaia} data release \citep[\textit{Gaia} DR2;][]{GaiaDR2}, and conducted follow-up observations of candidates within 200 pc. \citet{Benitez2025} expanded on this sample by following up on southern hemisphere candidates within 500\,pc. They identified 184 RGB stars, 75\,\% of which have a high probability of belonging to binary systems with orbital periods of up to 900 days. None of these were identified as double-lined spectroscopic binaries (SB2s) in the \textit{Gaia} DR3 non-single stars catalogue \citep[][]{GaiaDR3}.} Although their search was not specifically geared towards detecting SB2s, the absence of such systems in their follow-up spectroscopy is still notable. 
\citet{60containinggiants} compared binary evolution models to 60 medium- to long-period binaries. Fewer than ten of their systems include two giants on the first giant branch.
Finally, large spectroscopic surveys identified thousands of candidate SB2 systems, but only a few are within the red giant selection of \citet{Uzundag2022}: about 20 to 30 in the GALAH survey \citep{GALAH} and 13 in the APOGEE survey \citep{APOGEE}. 

In this work we present and analyse a rare instance of a twin red giant -- the SB2 \BD\ discovered during a binary red giant survey. It is composed of two giant stars with nearly identical masses and luminosities. This rare configuration provides an excellent opportunity to study binary interactions during the RGB phase.

We aimed to characterise the observational parameters of \BD\ and constrain possible outcomes of its post-interaction evolution, providing a benchmark case for the study of binary interaction in stellar evolution.
Section \ref{obs} details the observational data and spectra for this system. Section \ref{analysis} outlines our analysis of the radial velocity curve, spectral fitting, and evolutionary tracks in the Hertzsprung-Russell diagram \highlight{(HRD)}. In Sect. \ref{results} we explore potential evolutionary scenarios for this system, including possible binary evolution outcomes. Finally, Sect. \ref{conclusion} summarises our findings and discusses the prospects for future research.

\section{Target selection and observations}\label{obs}

\begin{figure}
   \centering
   \includegraphics[width=0.49\textwidth]{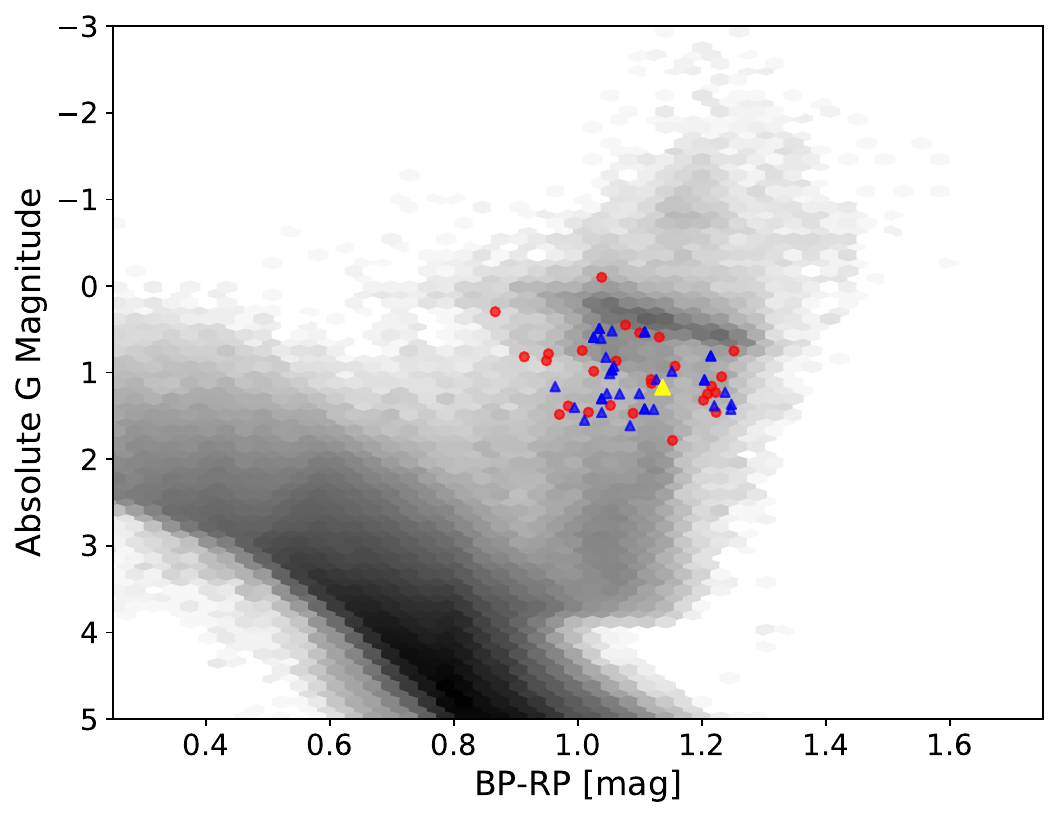}
      \caption{Colour-magnitude diagram of observed red giant stars, based on \textit{Gaia} DR3 \citep{GaiaDR3}. The sample includes RGB stars just below the red clump within 500 pc of the Sun. Blue triangles mark spectra from OES, red circles from AISAS. BD+20 5391 is highlighted in yellow. The grey background marks all \textit{Gaia} DR3 objects with \texttt{parallax\_over\_error} $>100$ within 500 pc of the Sun.}
         \label{RGsample}
\end{figure}

\BD\ was initially identified as a double-lined system during a search for RGB stars, aimed at finding progenitors of hot subdwarf and He-WD stars, complementing the independent studies of \cite{Uzundag2022} and \citet{Benitez2025}. Originally using \textit{Gaia} DR2, we constructed a volume-limited sample of RGB stars just below the red clump within 500 pc of the Sun (see Fig. \ref{RGsample}), thus excluding helium-burning giants and mirroring the selection of the 500 pc volume-limited hot subdwarf sample by \citet{Dawson2024}.
Out of a randomly selected sample of 100 stars located in the northern hemisphere and observable from the Ond\v{r}ejov Observatory and the Astronomical Institute of the Slovak Academy of Sciences (AISAS) at Skalnaté Pleso, 51 were observed using the high-resolution Ondřejov Echelle Spectrograph \citep[OES;][]{OES, Kabath2020} mounted on the Perek 2 m Telescope at Ondřejov Observatory, which has a maximum resolving power of \(R = 51\,600\) at 5000\,\AA\ and a spectral coverage between about 3800 and 8800\,\AA. The remaining 49 were observed at Skalnaté Pleso Observatory with the 1.3 m telescope and the MUSICOS-style \'echelle spectrograph \citep{Baudrand1992, Pribulla2015}. \BD\ is the only double-lined system identified in this sample and the focus of this work.

Spectroscopic observations of \BD\ were conducted from 2020 to 2025 using the OES, resulting in a total of 15 spectra. Our OES spectra feature an average mean signal-to-noise ratio per pixel of 30 to 35. 

We reduced the OES spectra using an Image Reduction and Analysis Facility \citep[IRAF;][]{iraf, Tody1993} package and a dedicated semi-automatic pipeline\footnote{\url{https://doi.org/10.5281/zenodo.8420746}} \citep{Cabezas_OESRED}, which incorporates a full range of standard procedures for \'echelle spectra reduction: bias correction, flat-fielding, wavelength calibration, heliocentric velocity correction, and continuum normalisation. 
Additionally, one spectrum was obtained in late 2024 with the High-Efficiency and high-Resolution Mercator Échelle Spectrograph \citep[HERMES;][]{HERMES} at the 1.2 m Mercator Telescope on La Palma, which offers a resolving power of \(R = 85\,000\) and a wavelength range from 3800 to 9000 $\mathrm{\AA}$, at a median signal-to-noise ratio of 59. The integrated HERMES data reduction pipeline performs standard corrections, robust spectral order tracing and extraction, pre-normalisation, and resampling or merging of spectra described in detail in \citet[][]{HERMES}.

\section{Analysis}\label{analysis}

Our analysis comprises four complementary methods: spectral analysis, radial velocity modelling, spectral energy distribution (SED) fitting, and HRD fitting to constrain the evolutionary state and future evolution of the system. The details of these methods are outlined in the following sections.
Our results are summarised in Table~\ref{tab:parameters}.  

\subsection{Radial velocity curve}\label{sectionRVcurve}

As a first step, we used the Interactive Spectral Interpretation System \citep[ISIS;][]{ISIS} to perform $\chi^2$ fits to all spectra. Composite spectral templates were employed to derive the radial velocities of both components in the double-lined spectrum of BD+20~5391. \highlight{The fitting procedure and model spectra are described in more detail in Sect. \ref{sectionspectral}.} Table~\ref{tab:rv_measurements} gives an overview of the radial velocities computed from the spectra of \BD.

\begin{figure}
   \centering
   \includegraphics[width=0.49\textwidth]{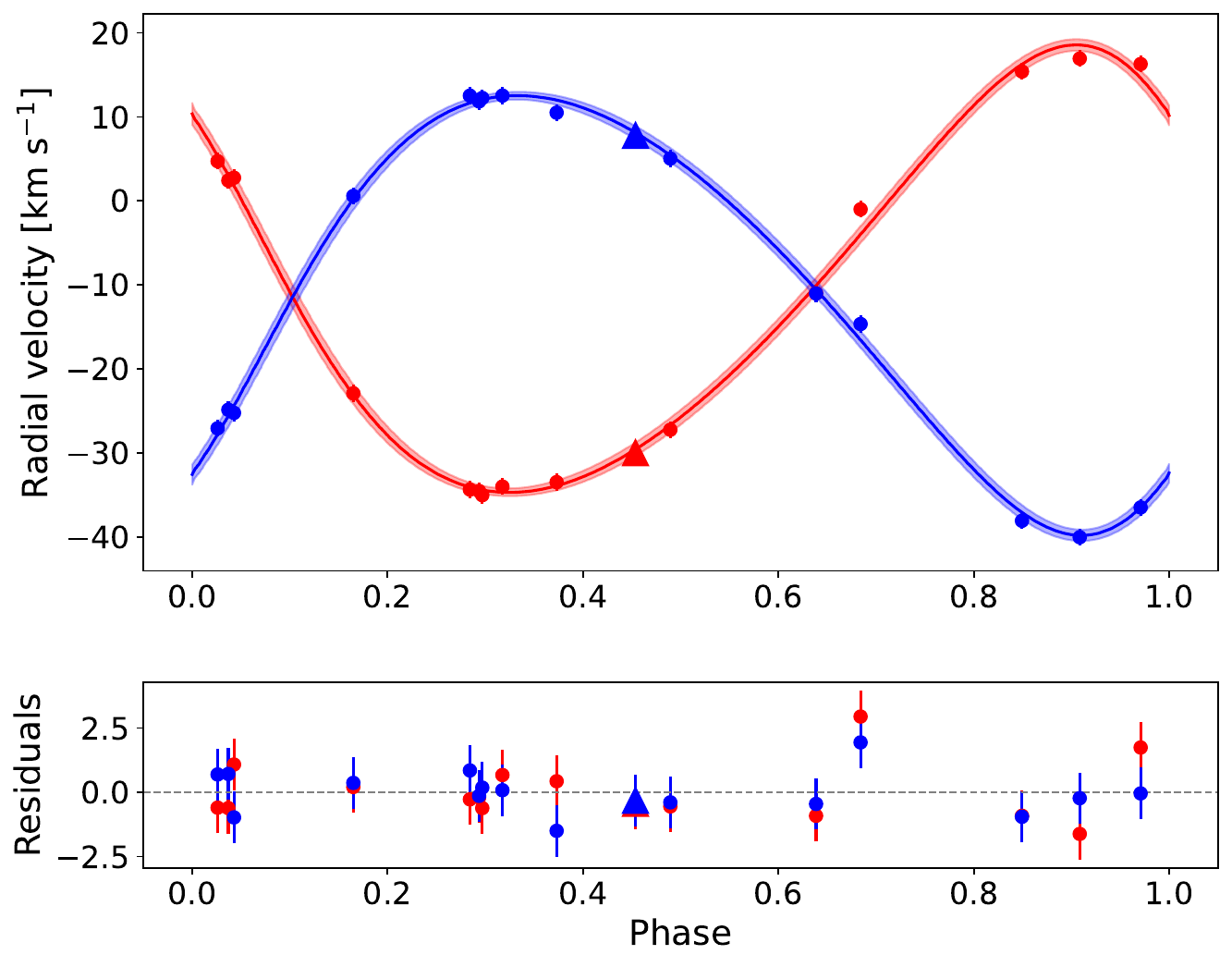}
      \caption{Radial velocity curve of BD+20~5391 for components A (red) and B (blue). 
      Circles show measurements obtained with the OES, and the triangles mark the HERMES measurement.}\label{rvcurve_plot}
\end{figure}

The radial velocities were then modelled assuming an eccentric orbit and fitted using a Markov chain Monte Carlo (MCMC) approach, implemented with the \texttt{emcee} package \citep{mcmc2013} in Python. 
The resulting radial velocity curves are presented in Fig.~\ref{rvcurve_plot}.
We defined the mass ratio as \(q=M_{\mathrm{A}}/M_\mathrm{B}\), where \(M_\mathrm{B}\) refers to the primary component, and \(M_\mathrm{A}\) to the secondary component, with \(M_\mathrm{B} > M_\mathrm{A}\). 
 
The orbital period, \(P = 80.9 \pm 0.5\) d, and radial velocity semi-amplitudes \(K_\mathrm{A}\) and \(K_\mathrm{B}\) (see Table~\ref{tab:parameters}) are in good agreement with the values reported in the \textit{Gaia} DR3 non-single star catalogue's  solution  \citep[][]{GaiaDR3}: \(P_\mathrm{\textit{Gaia}} = 81.054 \pm 0.028\) d, \(K_{\mathrm{A}, \mathrm{\textit{Gaia}}} = 27.7 \pm 0.7\) \kms, and \(K_{\mathrm{B}, \mathrm{\textit{Gaia}}} = 27.2 \pm 0.7\)\,\kms. However, the eccentricity derived from our spectra, \(e = 0.164 \pm 0.030\), is significantly lower than the \textit{Gaia} value of \(e_\mathrm{\textit{Gaia}} = 0.339 \pm 0.016\). Since we cannot find a well-matching solution with such a high eccentricity, we conclude that \textit{Gaia} likely overestimates the eccentricity of this system.

\subsection{Spectral analysis} \label{sectionspectral}

Given the higher resolution and well-defined continuum of the HERMES spectrum, we constrained the atmospheric parameters by performing a $\chi^2$ fit using only this spectrum.
For this, we computed a grid of \textsc{Atlas12}/\textsc{Synthe}  \citep[][]{Atlas,SYNTHE} models, which use the local thermal equilibrium (LTE) approximation in a 1D geometry. 
The grid covers a large parameter space in four dimensions: \teff\ from 3800 to 8000\,K in 200\,K steps, \logg\ from 5.6 to 1.4 in steps of 0.2, microturbulent velocity $\xi$ (0, 1, 2, 4\,\kms), and metallicity [Fe/H] from $-$2.5 to 0.5 in steps of 0.5.

Free parameters in the fit included the effective temperatures, radial velocities, and projected rotational velocities of both stars, as well as the surface ratio $S = R_{\mathrm{A}}^2/R_\mathrm{B}^2$. 
Given that the binary system likely formed as a coeval pair, we assumed a common scaled solar metallicity [Fe/H] for both components and no enhancement in alpha-process elements. 
We further used Newton's law of gravity $g = GM/R^2$ to impose an additional constraint on the surface gravities of the two components by expressing \(\log g_{\mathrm{A}}\) in terms of \(\log g_{\mathrm{B}}\): 
\begin{equation}
\log g_\mathrm{A} = \log g_\mathrm{B} + \log q - \log S,
\end{equation}
where $q=K_\mathrm{B}/K_\mathrm{A}=M_\mathrm{A}/M_\mathrm{B}$ is the mass ratio obtained from the radial velocity curve fit. 

\begin{figure*}
   \centering
   \includegraphics[width=\textwidth]{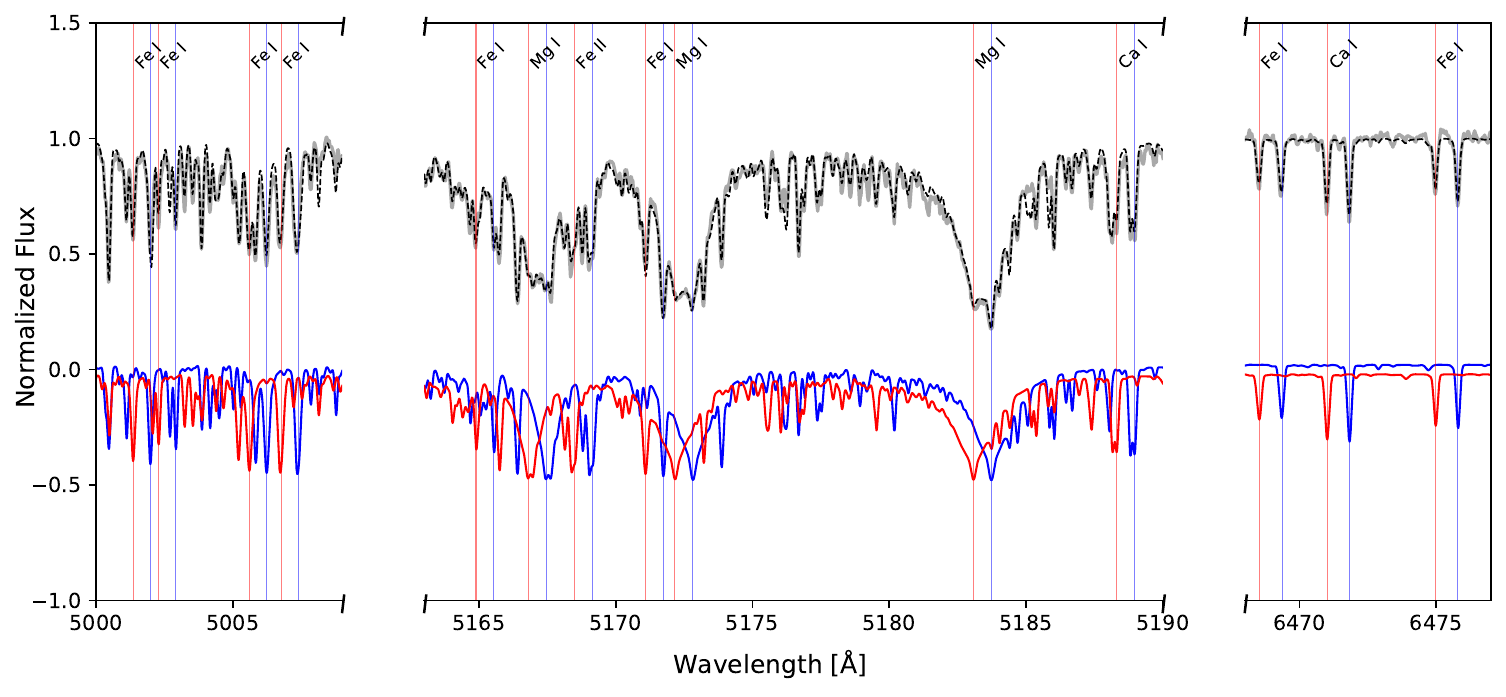}
      \caption{Selected ranges in the double-lined HERMES spectrum of \BD. 
      The upper section shows the observed data in light grey and the combined model fit as a dashed black line, while the lower part displays the contributions of the individual components, offset by $-$0.5 for clarity.
      Component A is shown in  red, component B in blue; strong lines are labelled. }
         \label{spectrum_plot}
\end{figure*}

An excerpt of our best-fit model fit is shown in Fig.~\ref{spectrum_plot}, highlighting the clearly resolved contributions of both stars. 
The derived atmospheric parameters are presented in Table~\ref{tab:parameters}.

To account for deficiencies in our model, such as the use of the LTE and 1D approximations, we assumed systematic uncertainties of \(\pm 50\) K to \teff\ and 0.1 to \(\log g\) for the parameters derived from the spectral fit, which is roughly consistent with the study of \citet[][Fig. 8]{Blanco2019} for grid-based model fits. For $\xi$ and $v\sin{i}$ we assumed uncertainties of 0.1 and 0.2, respectively, since the impact of statistical errors is negligible due to the high S/N and large number of data points in the HERMES spectrum. \highlight{We note that the binary nature of \BD\ may introduce additional uncertainties that are difficult to quantify. However, since both components are well resolved in the high-quality HERMES spectra, these uncertainties remain statistically small.}

\subsection{Spectral energy distribution}

To constrain the stellar parameters of both components, we combined the spectroscopic parameters with the \textit{Gaia} parallax \citep{Lindegren2021} and the observed SED, as constructed from photometric measurements. 
Our SED fitting method is described in detail by \citet{hebersed}. 

Here we performed a \(\chi^2\) minimisation to obtain the best-fit angular diameter ($\Theta$). The spectral fit parameters $\xi$, [Fe/H], \(\log g_\mathrm{A}\), \(\log g_\mathrm{B}\), \(T_{\mathrm{eff},\mathrm{A}}\), \(T_{\mathrm{eff},\mathrm{B}}\), and the surface ratio (\(S\)) were treated as fixed inputs.

In the fitting procedure, we accounted for reddening using the extinction law of  \citet{Fitzpatrick2019}, with an extinction parameter \(R_{55} = 3.02\), which is standard for the Galaxy's diffuse interstellar medium. To limit the number of free parameters, we fixed the monochromatic colour excess to a value of \(E(44-55)=0.06\) mag, based on the estimates provided by the \citet{SFD1998}, \citet{SF2011}, \citet{Capitanio2017}, and \citet{Green2019} reddening maps.
Our model predicts a marginally higher near-ultraviolet flux compared to the observed values from \textit{Gaia}, which may be caused by limitations in the \textsc{Synthe} model used for the spectral fit.

To obtain the stellar radius,
\begin{equation}
    R = \Theta/(2\varpi),
\end{equation}
the angular diameter was then combined with the parallax \(\varpi = 2.341 \pm 0.019\) mas provided by Data Release 3 of the \textit{Gaia} mission \citep{GaiaDR3}. We corrected the parallax for its zero-point offset following \citet{Lindegren2021} and inflated the corresponding uncertainty using the function suggested by \citet{El-Badry2021}.
The luminosity \(L\) was calculated using
\begin{equation}
    L = (R/R_\odot)^2(T_{\mathrm{eff}}/T_{\mathrm{eff},\odot})^4\; .
\end{equation}

\begin{figure}
   \centering
   \includegraphics[width=0.49\textwidth]{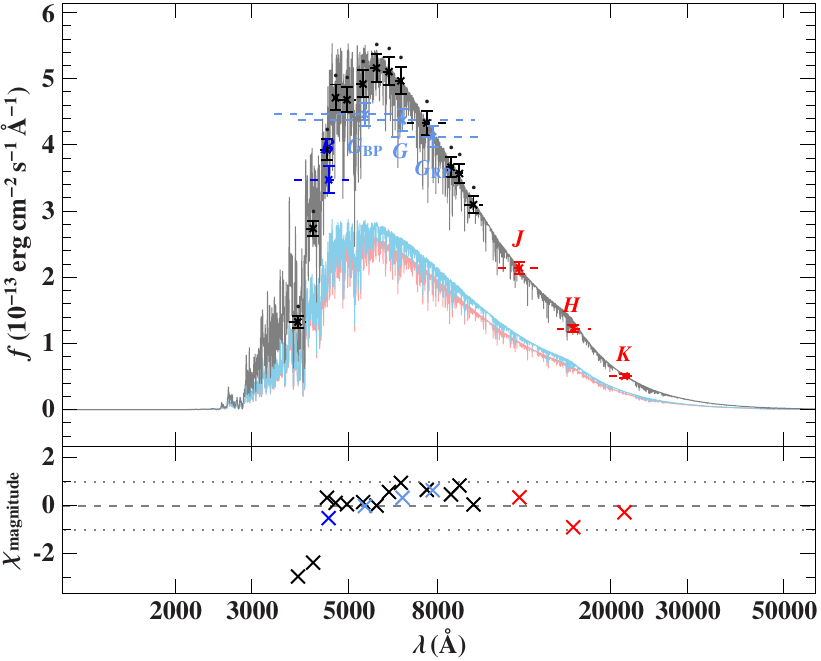}
      \caption{SED of BD+20~5391. Component A is red, component B blue, and the combined model is grey. Photometric measurements are provided by \textit{Gaia}/XP \citep[black;][]{DeAngeli2023_XP}, \textit{Gaia} \citep[light blue;][]{Riello2021}, APASS \citep[blue;][]{Henden_apass9}, and 2MASS \citep[red;][]{Skrutskie2006_2MASS}. Interstellar extinction was applied to the models. }
         \label{SED_plot}
\end{figure}

Table~\ref{tab:parameters} lists the resulting parameters. Figure \ref{SED_plot} shows the SED. Using the spectroscopic surface gravity and Newton’s law of gravity, we derived a stellar mass of \(M_\mathrm{A}=1.5 \pm 0.4\)\,\msun\ for both stars. 
The large uncertainty on these masses originates from the limited accuracy of our spectroscopic surface gravity estimates (about 0.1\,dex). 

To investigate potential photometric variability due to pulsations or rotational modulation, we checked the light curve obtained from the Transiting Exoplanet Survey Satellite \citep[TESS;][]{TESS_2015}, in particular the reduced full-frame light curves from Sector 83, available through the Barbara Mikulski Archive for Space Telescopes (MAST\footnote{\url{https://mast.stsci.edu/portal/Mashup/Clients/Mast/Portal.html}}). However, no significant variability was detected. Even without photometric data to constrain the rotation periods, the $v\sin i$ values of both components are similarly low and close to the spectral resolution, which suggests comparable rotation rates.

\subsection{Evolution and onset of Roche lobe overflow}

Leveraging the parameters derived from the SED fitting, we constrained the position of \BD\ in the HRD. To further characterise the evolutionary state of both stellar components, we applied an interpolation approach based on MESA Isochrone and Stellar Tracks Collaboration \citep[MIST;][]{Choi_2016} evolutionary tracks; this enabled a refined determination of their properties.

The tracks were interpolated using the \texttt{scipy} Python module \citep{2020Scipy} to create a multi-dimensional grid, enabling a smooth interpolation across initial mass, metallicity, and rotation for each equivalent evolutionary point. This interpolation function returns stellar parameters such as effective temperature (\(\log T_\mathrm{eff}\)), luminosity (\(\log L\)), mass, radius, and age at a given equivalent evolutionary point.

The observed stellar parameters, effective temperature and luminosity, are compared to the interpolated MIST tracks using an MCMC approach with Bayesian constraints. 
We then used the \texttt{emcee} Python module to sample the posterior distribution of the stellar parameters. In this process we minimised the likelihood function, 
\begin{equation}
\log \mathcal{L} = -\frac{1}{2} \left( \chi_1^2 + \chi_2^2 \right),
\end{equation}
where the individual chi-squared values for each star were computed based on the observed effective temperature (\(\log T_\mathrm{eff}\)) and luminosity (\(\log L\)) compared to the interpolated values at a given evolutionary phase, as
\begin{equation}
\chi^2 = \left( \frac{\log L - \log L_\mathrm{obs}}{\sigma_{\log L}} \right)^2 + \left( \frac{\log T_\mathrm{eff} - \log T_\mathrm{eff, obs}}{\sigma_{\log T}} \right)^2.
\end{equation}
To constrain the parameter space, we incorporated Gaussian prior distributions for metallicity based on the spectroscopic analysis (\(\mathrm{[Fe/H]} = -0.136 \pm 0.020\)) and for the mass ratio, as derived from the radial velocity curve (\( q = 0.981 \pm 0.021\)). The resulting posterior distributions for the fitted parameters are shown in the top corner plot of Fig.~\ref{fig:corner_plots}, and the resulting parameters are listed in Table \ref{tab:parameters}.

The current semi-major axis of the binary orbit based on our fit is
\begin{equation}
    a = a_\mathrm{A}+a_\mathrm{B} = \left( \frac{P^2 G (M_\mathrm{A} + M_\mathrm{B})}{4\pi^2} \right)^{1/3}= 110.8^{+2.5}_{-2.4} \, R_{\odot}
.\end{equation}
At an orbital period of only 81\,days, and with a separation at periastron of \(r_{\mathrm{peri}} = a(1 - e) = 92.6^{+2.1}_{-2.0} \, R_{\odot} \), the components of BD+20~5391 will begin mass transfer before they reach the tip of the RGB. We estimated the onset of RLOF by comparing stellar and Roche lobe radii along MIST tracks, iterating over MCMC samples to derive the corresponding age and helium core mass distribution.
The Roche lobe radius was calculated using the \citet{EggletonRL} formula:\\
\begin{equation}
r_\mathrm{L} = \frac{0.49 q^{2/3}}{0.6 q^{2/3} + \log(1 + q^{1/3})},
\end{equation}
where \(r_\mathrm{L}\) is the Roche lobe radius in units of separation. 

The onset of RLOF was determined by comparing the stellar radius with the corresponding Roche lobe radius throughout the evolutionary track. If the stellar radius exceeds the Roche lobe radius, the star is considered to undergo RLOF. The onset of RLOF is considered to be the point where the first of the two stars fills out its Roche lobe. 

At the onset of RLOF, we recorded the age, stellar radii, and helium core masses. This process was repeated for all posterior samples from the MCMC calculation, generating a distribution of RLOF ages and helium core masses. These distributions, along with the parameters derived from the fitting of the evolutionary tracks, are presented in the bottom plot of Fig. \ref{fig:corner_plots}. The corresponding values are listed in Table~\ref{tab:parameters}. 

We performed a similar analysis using the BaSTI \citep{BaSTI2021} stellar evolution models. While the resulting masses and radii are in good agreement with those previously discussed, the inferred metallicity from the HRD fit is significantly higher ([Fe/H] = +0.08) and inconsistent with the rest of our analysis. We therefore proceeded with the results based on the MIST tracks, noting that there is an intrinsic uncertainty to such model tracks.

\begin{table}
    \setstretch{1.15}
    \caption{System- and component-specific parameters.}
    \label{tab:parameters}
    \centering
    \scalebox{0.9}{
    \begin{tabular}{l c c}
        \hline
        \hline
        Parameter & Component A & Component B \\
        \hline
        \multicolumn{3}{c}{Radial velocity analysis} \\
        \hline
        $P$ (d) & \multicolumn{2}{c}{$80.9 \pm 0.5$} \\
        $e$ & \multicolumn{2}{c}{$0.164 \pm 0.030$} \\
        $q=M_\mathrm{A}/M_\mathrm{B}$ & \multicolumn{2}{c}{$0.981 \pm 0.021$} \\
        $K$ (km\,s$^{-1}$) & $26.7 \pm 0.5$ & $26.2 \pm 0.4$ \\
        \hline
        \multicolumn{3}{c}{Spectral analysis} \\
        \hline
        $[$Fe/H$]$ & \multicolumn{2}{c}{$-0.136 \pm 0.020$} \\
        $S = R_\mathrm{A}^2/R_\mathrm{B}^2$ & \multicolumn{2}{c}{$0.9 \pm 0.05$} \\
        $T_{\mathrm{eff}}$ (K) & $4934 \pm 50$ & $4913 \pm 50$ \\
        $\log g$ (cgs) & $3.18 \pm 0.10$ & $3.14 \pm 0.10$ \\
        $\xi$ (km\,s$^{-1}$) & $1.0 \pm 0.1$ & $1.3 \pm 0.1$\\
        $v\sin{i}$ (km\,s$^{-1}$) & $4.1 \pm 0.2$ & $3.5 \pm 0.2$\\
        \hline
        \multicolumn{3}{c}{SED analysis} \\
        \hline
        $E$\,(44-55) (mag) & \multicolumn{2}{c}{$0.06$ (fixed)}\\
        $\log \Theta \,(\mathrm{rad}) $ & $-9.265 \pm 0.015$ & $-9.242 \pm 0.015$\\
        $L$ (\lsun) & $14.2 \pm 1.5$ & $15.5 \pm 1.3$ \\
        $R$ (\rsun) & $5.15 \pm 0.23$ & $5.43 \pm 0.19$ \\
        $M$ (\msun) & $1.5 \pm 0.5$ & $1.5 \pm 0.5$ \\
        \hline
        \multicolumn{3}{c}{HRD fit} \\
        \hline
        $M$ (\msun) & $1.38 \pm 0.10$ & $1.41 \pm 0.10$\\
        $\log\tau$ (yr) & $9.50 \pm 0.10$ & $9.47\pm 0.10$ \\
        \hline
        \multicolumn{3}{c}{At the onset of RLOF} \\
        \hline
        $\log\tau_\mathrm{RLOF}$ (yr) & $9.52 \pm 0.10$ & $9.49\pm 0.10$ \\
        $R$ (\rsun) & $34.9 \pm 0.8$ & $34.8 \pm 0.8$ \\
        $M_\mathrm{He}$ (\msun) & $0.335 \pm 0.003$ & $0.335 \pm 0.003$ \\
        \hline
    \end{tabular}}
\end{table}

\section{Results and discussion}\label{results}

\begin{figure*}
   \centering
   \includegraphics[width=0.49\textwidth]{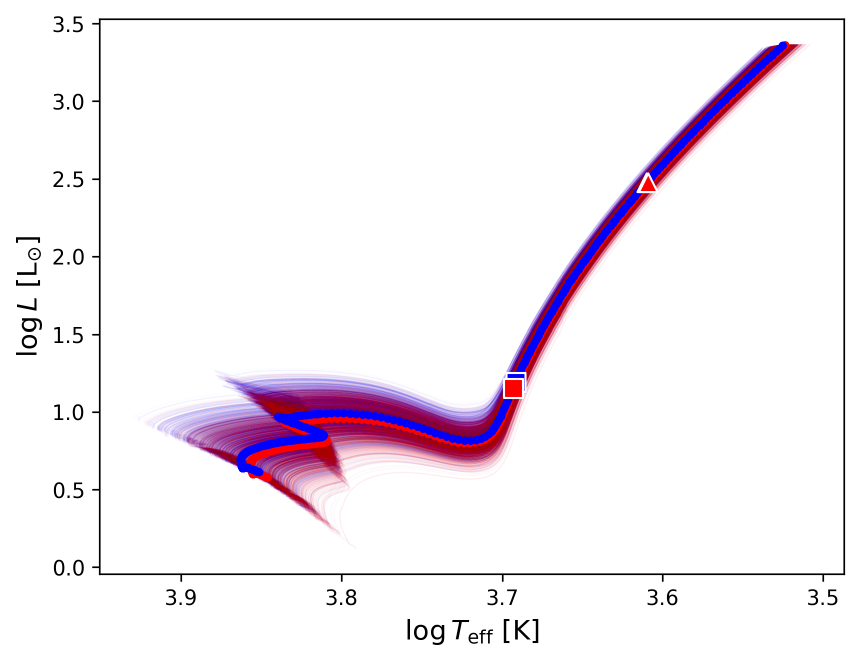}
   \includegraphics[width=0.49\textwidth]{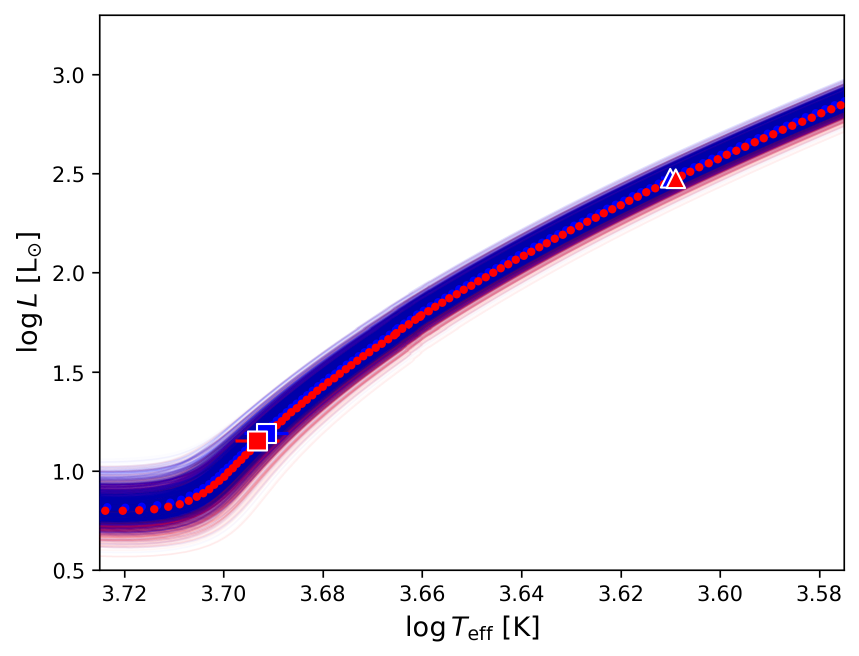}
   \caption{HRD for BD+20~5391. Component A is marked in red and component B in blue. The squares denote the parameters derived from the SED and spectral fitting, while triangles indicate the onset of RLOF. The best-fit MIST evolutionary tracks are represented by discrete dots extending from the zero-age main sequence to the tip of the RGB. To visualise the uncertainties, we also show a random subset of tracks from our MCMC calculation.}
   \label{fig:hrd_plots}
\end{figure*}

The MCMC techniques applied to model the evolutionary state of BD+20~5391 produced corner plots (Fig.\ \ref{fig:corner_plots}) and HRDs (Fig.\,~\ref{fig:hrd_plots}), providing constraints on the stellar parameters. Our results indicate that RLOF is set to occur for both components before they reach the tip of the RGB. 
Based on our MCMC results, the two stars are expected to fill their Roche lobes at essentially the same time. 
The time remaining until RLOF for each component is calculated as \(\Delta\tau = \tau_\mathrm{RLOF} - \tau\), yielding \(\log \Delta\tau_\mathrm{A} = 8.20 \pm 0.04\,\mathrm{yr}\) and \(\log \Delta\tau_\mathrm{B} = 8.16 \pm 0.04\,\mathrm{yr}\). The delay between the two RLOF events is then given by \(\Delta t_\mathrm{RLOF} = (\Delta\tau)_\mathrm{A} - (\Delta\tau)_\mathrm{B} = 13 \pm 15\,\mathrm{Myr}\), which is consistent with simultaneous RLOF.
Since the Eggleton formula provides an upper limit for the Roche lobe radius, the onset of mass transfer is likely to begin earlier due to additional factors such as stellar wind mass loss and the gradual expansion of the stellar atmospheres, further supporting a nearly simultaneous onset of RLOF for the two stars. This makes BD+20~5391 a strong candidate for undergoing double-core CE evolution.

The most notable outcome of our analysis is that the helium core masses at the onset of RLOF are essentially identical: at that moment, components A and B are both predicted to have a core mass of $0.335 \pm 0.003$\,\msun\ (statistical uncertainty only). 
Two primary evolutionary pathways emerge from our analysis: one in which the two stars merge and another in which they remain separate.

\subsection{Merger scenario}
\label{sect:merger}

If the binary components merge following RLOF, the combined helium core mass would be approximately 0.7\,$M_{\odot}$, which is sufficient for the remnant to ignite helium fusion. 
If the envelope is not ejected, this would result in a relatively massive red clump star. 
However, if the remaining envelope is expelled -- which depends on the poorly constrained CE ejection efficiency -- the two He cores might end up in a binary close enough that angular momentum loss through the emission of gravitational waves might lead to orbital shrinkage and an eventual merger. The remnant would likely turn into a relatively massive, core helium-burning hot subdwarf star.
In particular, most helium-rich sdO (He-sdO) stars are thought to form from double He-WD mergers \citep{Webbink1984, Zhang2012, Schwab2018, Yu2021}.
Approximately one percent of He-sdOs exhibit strong magnetic fields, with average strengths between 200 and 500 kG \citep{Matti2022, Dorsch2024, Pelisoli2022}. Notably, all of these magnetic He-sdOs have observed masses of about 0.8\,\msun, which is close to the maximum mass attainable through He-WD mergers. This mass constraint implies that the two progenitor He-WDs must have been coeval, as expected for \BD.
Indeed, \cite{Pakmor2024} recently suggested that magnetic He-sdOs originate from the merger of two equal-mass He-WDs, based on 3D magnetohydrodynamic simulations.

\subsection{Non-merger scenario}

If the system is able to eject a CE, but avoids merging, the result would be a double low-mass He-WD binary, likely with a short orbital period. 
Observational evidence of systems potentially formed through this evolutionary pathway has recently been reported by \cite{Munday2024}, who identified approximately a dozen binaries with comparable component masses in the range of 0.2 to 0.4\,\msun, and by \cite{Burdge2020}.

\citet{Justham_HerichsdO} argue speculatively that systems consisting of two helium-rich hot subdwarfs could be the outcome of a double core CE phase as might happen for BD+20~5391. 
This includes the unique He-sdB binary PG~1544+488\footnote{HE~0301-3039 was originally thought to be a PG~1544+488-like system \citep{HE03} but turned out to be single-lined and shows no radial velocity variations.} \citep{PG15disc, Sener2014}.

The orbital separation after the interaction, which depends on the efficiency of mass transfer and envelope loss, will determine whether the system remains a wide double He-WD or undergoes further interactions. If the final separation is large, the system could persist as a stable double-degenerate system for a Hubble time.

\section{Conclusions and outlook}\label{conclusion}

Our analysis of \BD\ reveals it to be a rare SB2 consisting of two nearly identical 1.4\,\msun\ red giant stars in an 81-day orbit. The system’s orbital and stellar parameters suggest that the two stars have evolved in parallel without any significant mass transfer thus far.

By approximating their evolution using single-star MIST tracks, we find that the two stars will likely fill their Roche lobes almost simultaneously. This could lead to the so-called double-core evolution scenario, where the two stars orbit within a CE. Predicting their subsequent evolution requires detailed modelling that accounts for mass transfer, tidally enhanced stellar winds \citep{Tout1988,Han1998}, and the complex physics of the CE phase \citep{Taam2010,Ivanova2013}.

In our simplified model and given our choice of MIST tracks, both stars will have degenerate helium cores of 0.33\,\msun\ when RLOF begins. If the envelope is ejected without merging, the system can evolve into a close double He-WD binary. Alternatively, the stars could merge, forming a possibly magnetic 0.7\,\msun\ He-sdO, a type of star that is known to exist \citep{Dorsch2024, Pelisoli2022, Pakmor2024}.  

BD+20~5391 thus provides a valuable observational benchmark for studying the progenitors of evolved compact objects such as hot subdwarfs and double He-WDs. We encourage future detailed evolutionary studies of this system. 
In addition, large spectroscopic surveys such as {\it Gaia} DR4+ and 4MOST \citep{4MOST_2019} will not only identify more systems like BD+20~5391 but also detect their possible descendants -- double He-WDs and He-sdO stars. By observing systems before and after mass transfer, these surveys will help refine binary population synthesis models and improve our understanding of low- and intermediate-mass binary evolution. This will build upon previous binary population synthesis studies, such as those by \cite{Han2003} for hot subdwarfs and \cite{Nelemans2005} and \cite{vanderSluys2006} for white dwarfs.

With the release of DR3, \textit{Gaia}’s Radial Velocity Spectrometer already enables the detection of some binaries and the determination of their spectroscopic orbits \citep{GaiaDR3_binary}. 
Within the colour-magnitude selection of \citet{Uzundag2022}, 19 SB2 red giants, including \BD, meet the criteria of having orbital periods between 20 and 250 days, mass ratios ($q$) > $0.95$, and eccentricities ($e$) $<0.9$. 
Future careful analyses of these systems are necessary to check whether they are genuine double RGB stars since this is not always the case: RZ Eri had  previously been identified as an RGB + main sequence system \citep{60containinggiants}, while BD+49\,1747 and HD\,7731 are likely red clump stars \citep{Ruiz-Dern2018}.

\begin{acknowledgements}

We thank the anonymous referee for their helpful suggestions, which have improved the quality of this paper. Research is based on data taken with the Perek telescope at the Astronomical Institute of the Czech Academy of Sciences (ASU) in Ondřejov. We thank the Stellar Physics Department of ASU and their technical staff for the support during the observations. This research was supported by funds from DAAD PPP Czech Republic project 57509654 and DAAD PPP Slovakia project 57513233. BK acknowledge the support from the Mobility Plus Project DAAD-24-02 and RVO:67985815. We thank the participants of the 2021 Ond\v{r}ejov-Potsdam workshop for their observing assistance: Ayesha Arshad Arain, Saksham Arora, Radha Anil Gharapurkar, Siddarth Khalate, Chinmay Mahajan, Henrik Rose, Amrit Sedain, Tahereh Ramezani, Caiun Xia, Prapti Mondal, and Kate\v{r}ina Pivo\v{n}kov\'{a}.
MD was supported by the Deutsches Zentrum für Luft- und Raumfahrt (DLR) through grant 50-OR-2304. 
VS and MP were supported by the Deutsche Forschungsgemeinschaft (DFG) through grants GE2506/9-1 and GE2506/12-1, MP additionally through grant GE2506/18-1. IP acknowledges support by the DFG through grant GE2506/12-1 and from the Royal Society through a University Research Fellowship (URF/R1/231496). HD was supported by the DFG through grants GE2506/17-1 and GE2506/9-2. 
EK and JB acknowledge VEGA 2/0031/22. 
This research has used measurements obtained at the Mercator Observatory which receives funding from the Research Foundation Flanders (FWO) (grant agreement I000325N and I000521N). KD acknowledges funding from grant METH/24/012 at KU Leuven.
We thank Rainer Hainich and Fabian Mattig for maintaining the Astronomical Observation Tracking System (AOTS) database. 
EK acknowledges support from the Slovak Research and Development Agency under contract No. APVV-20-0148.
This work has made use of data from the European Space Agency (ESA) mission {\it Gaia} (\url{https://www.cosmos.esa.int/gaia}), processed by the {\it Gaia} Data Processing and Analysis Consortium (DPAC, \url{https://www.cosmos.esa.int/web/gaia/dpac/consortium}). Funding for the DPAC has been provided by national institutions, in particular the institutions participating in the {\it Gaia} Multilateral Agreement. 
This paper includes data collected with the TESS mission, obtained from the MAST data archive at the Space Telescope Science Institute (STScI). Funding for the TESS mission is provided by the NASA Explorer Program. STScI is operated by the Association of Universities for Research in Astronomy, Inc., under NASA contract NAS 5–26555. 
IRAF is distributed by the National Optical Astronomy Observatories, operated by the Association of Universities for Research in Astronomy, Inc. under a cooperative agreement with the National Science Foundation. 
This research has made use of NASA's Astrophysics Data System.

\end{acknowledgements}

\bibliographystyle{aa} 
\bibliography{aa56201-25}

\hyphenation{Post-Script Sprin-ger}
\begin{thebibliography}{81}
\expandafter\ifx\csname natexlab\endcsname\relax\def\natexlab#1{#1}\fi

\bibitem[{{Ahmad} {et~al.}(2004){Ahmad}, {Jeffery}, \& {Fullerton}}]{PG15disc}
{Ahmad}, A., {Jeffery}, C.~S., \& {Fullerton}, A.~W. 2004, \aap, 418, 275

\bibitem[{{Baudrand} \& {Bohm}(1992)}]{Baudrand1992}
{Baudrand}, J. \& {Bohm}, T. 1992, \aap, 259, 711

\bibitem[{{Beck} {et~al.}(2024){Beck}, {Grossmann}, {Steinwender}, {Schimak},
  {Muntean}, {Vrard}, {Patton}, {Merc}, {Mathur}, {Garcia}, {Pinsonneault},
  {Rowan}, {Gaulme}, {Allende Prieto}, {Arellano-C{\'o}rdova}, {Cao},
  {Corsaro}, {Creevey}, {Hambleton}, {Hanslmeier}, {Holl}, {Johnson}, {Mathis},
  {Godoy-Rivera}, {S{\'\i}mon-D{\'\i}az}, \& {Zinn}}]{Beck2024}
{Beck}, P.~G., {Grossmann}, D.~H., {Steinwender}, L., {et~al.} 2024, \aap, 682,
  A7

\bibitem[{{Beck} {et~al.}(2018){Beck}, {Kallinger}, {Pavlovski}, {Palacios},
  {Tkachenko}, {Mathis}, {Garc{\'\i}a}, {Corsaro}, {Johnston}, {Mosser},
  {Ceillier}, {do Nascimento}, \& {Raskin}}]{Beck2018}
{Beck}, P.~G., {Kallinger}, T., {Pavlovski}, K., {et~al.} 2018, \aap, 612, A22

\bibitem[{{Beck} {et~al.}(2022){Beck}, {Mathur}, {Hambleton}, {Garc{\'\i}a},
  {Steinwender}, {Eisner}, {do Nascimento}, {Gaulme}, \& {Mathis}}]{Beck2022}
{Beck}, P.~G., {Mathur}, S., {Hambleton}, K., {et~al.} 2022, \aap, 667, A31

\bibitem[{Belczyński \& Kalogera(2001)}]{Belczy_ski2001}
Belczyński, K. \& Kalogera, V. 2001, The Astrophysical Journal, 550,
  L183–L187

\bibitem[{{Benitez-Palacios} {et~al.}(2025){Benitez-Palacios}, {Uzundag},
  {Vu{\v{c}}kovi{\'c}}, {Arancibia-Rojas}, {Dur{\'a}n-Reyes}, {Vos}, {Bobrick},
  {Zorotovic}, \& {Jones}}]{Benitez2025}
{Benitez-Palacios}, D., {Uzundag}, M., {Vu{\v{c}}kovi{\'c}}, M., {et~al.} 2025,
  \aap, 697, A98

\bibitem[{{Blanco-Cuaresma}(2019)}]{Blanco2019}
{Blanco-Cuaresma}, S. 2019, \mnras, 486, 2075

\bibitem[{{Brogaard} {et~al.}(2022){Brogaard}, {Arentoft}, {Slumstrup},
  {Grundahl}, {Lund}, {Arndt}, {Grund}, {Rudrasingam}, {Theil}, {Christensen},
  {Sejersen}, {Vorgod}, {Salmonsen}, {{\O}rtoft Endelt}, {Dainese}, {Frandsen},
  {Miglio}, {Tayar}, \& {Huber}}]{Brogaard2022}
{Brogaard}, K., {Arentoft}, T., {Slumstrup}, D., {et~al.} 2022, \aap, 668, A82

\bibitem[{{Brown}(1995)}]{Brown1995}
{Brown}, G.~E. 1995, \apj, 440, 270

\bibitem[{{Burdge} {et~al.}(2020){Burdge}, {Prince}, {Fuller}, {Kaplan},
  {Marsh}, {Tremblay}, {Zhuang}, {Bellm}, {Caiazzo}, {Coughlin}, {Dhillon},
  {Gaensicke}, {Rodr{\'\i}guez-Gil}, {Graham}, {Hermes}, {Kupfer},
  {Littlefair}, {Mr{\'o}z}, {Phinney}, {van Roestel}, {Yao}, {Dekany}, {Drake},
  {Duev}, {Hale}, {Feeney}, {Helou}, {Kaye}, {Mahabal}, {Masci}, {Riddle},
  {Smith}, {Soumagnac}, \& {Kulkarni}}]{Burdge2020}
{Burdge}, K.~B., {Prince}, T.~A., {Fuller}, J., {et~al.} 2020, \apj, 905, 32

\bibitem[{{Cabezas} {et~al.}(2023){Cabezas}, {{\v{S}}lechta}, {{\v{S}}koda}, \&
  {Kub{\'a}tov{\'a}}}]{Cabezas_OESRED}
{Cabezas}, M., {{\v{S}}lechta}, M., {{\v{S}}koda}, P., \& {Kub{\'a}tov{\'a}},
  B. 2023, {OESRED, the semi-automatic reduction code for Ond{\v{r}}ejov
  Echelle Spectrograph}

\bibitem[{{Capitanio} {et~al.}(2017){Capitanio}, {Lallement}, {Vergely},
  {Elyajouri}, \& {Monreal-Ibero}}]{Capitanio2017}
{Capitanio}, L., {Lallement}, R., {Vergely}, J.~L., {Elyajouri}, M., \&
  {Monreal-Ibero}, A. 2017, \aap, 606, A65

\bibitem[{Choi {et~al.}(2016)Choi, Dotter, Conroy, Cantiello, Paxton, \&
  Johnson}]{Choi_2016}
Choi, J., Dotter, A., Conroy, C., {et~al.} 2016, The Astrophysical Journal,
  823, 102

\bibitem[{{Clayton}(2012)}]{RCB2012}
{Clayton}, G.~C. 2012, Journal of the American Association of Variable Star
  Observers, 40, 539

\bibitem[{{{\c{S}}ener} \& {Jeffery}(2014)}]{Sener2014}
{{\c{S}}ener}, H.~T. \& {Jeffery}, C.~S. 2014, \mnras, 440, 2676

\bibitem[{{Dawson} {et~al.}(2024){Dawson}, {Geier}, {Heber}, {Pelisoli},
  {Dorsch}, {Schaffenroth}, {Reindl}, {Culpan}, {Pritzkuleit}, {Vos},
  {Soemitro}, {Roth}, {Schneider}, {Uzundag}, {Vu{\v{c}}kovi{\'c}}, {Antunes
  Amaral}, {Istrate}, {Justham}, {{\O}stensen}, {Telting}, {Djupvik}, {Raddi},
  {Green}, {Jeffery}, {Kepler}, {Munday}, {Steinmetz}, \&
  {Kupfer}}]{Dawson2024}
{Dawson}, H., {Geier}, S., {Heber}, U., {et~al.} 2024, \aap, 686, A25

\bibitem[{{De Angeli} {et~al.}(2023){De Angeli}, {Weiler}, {Montegriffo},
  {Evans}, {Riello}, {Andrae}, {Carrasco}, {Busso}, {Burgess}, {Cacciari},
  {Davidson}, {Harrison}, {Hodgkin}, {Jordi}, {Osborne}, {Pancino},
  {Altavilla}, {Barstow}, {Bailer-Jones}, {Bellazzini}, {Brown}, {Castellani},
  {Cowell}, {Delchambre}, {De Luise}, {Diener}, {Fabricius}, {Fouesneau},
  {Fr{\'e}mat}, {Gilmore}, {Giuffrida}, {Hambly}, {Hidalgo}, {Holland},
  {Kostrzewa-Rutkowska}, {van Leeuwen}, {Lobel}, {Marinoni}, {Miller},
  {Pagani}, {Palaversa}, {Piersimoni}, {Pulone}, {Ragaini}, {Rainer},
  {Richards}, {Rixon}, {Ruz-Mieres}, {Sanna}, {Sarro}, {Rowell}, {Sordo},
  {Walton}, \& {Yoldas}}]{DeAngeli2023_XP}
{De Angeli}, F., {Weiler}, M., {Montegriffo}, P., {et~al.} 2023, \aap, 674, A2

\bibitem[{{de Jong} {et~al.}(2019){de Jong}, {Agertz}, {Berbel}, {Aird},
  {Alexander}, {Amarsi}, {Anders}, {Andrae}, {Ansarinejad}, {Ansorge},
  {Antilogus}, {Anwand-Heerwart}, {Arentsen}, {Arnadottir}, {Asplund}, {Auger},
  {Azais}, {Baade}, {Baker}, {Baker}, {Balbinot}, {Baldry}, {Banerji},
  {Barden}, {Barklem}, {Barth{\'e}l{\'e}my-Mazot}, {Battistini}, {Bauer},
  {Bell}, {Bellido-Tirado}, {Bellstedt}, {Belokurov}, {Bensby}, {Bergemann},
  {Bestenlehner}, {Bielby}, {Bilicki}, {Blake}, {Bland-Hawthorn}, {Boeche},
  {Boland}, {Boller}, {Bongard}, {Bongiorno}, {Bonifacio}, {Boudon}, {Brooks},
  {Brown}, {Brown}, {Br{\"u}ggen}, {Brynnel}, {Brzeski}, {Buchert},
  {Buschkamp}, {Caffau}, {Caillier}, {Carrick}, {Casagrande}, {Case}, {Casey},
  {Cesarini}, {Cescutti}, {Chapuis}, {Chiappini}, {Childress}, {Christlieb},
  {Church}, {Cioni}, {Cluver}, {Colless}, {Collett}, {Comparat}, {Cooper},
  {Couch}, {Courbin}, {Croom}, {Croton}, {Daguis{\'e}}, {Dalton}, {Davies},
  {Davis}, {de Laverny}, {Deason}, {Dionies}, {Disseau}, {Doel}, {D{\"o}scher},
  {Driver}, {Dwelly}, {Eckert}, {Edge}, {Edvardsson}, {Youssoufi}, {Elhaddad},
  {Enke}, {Erfanianfar}, {Farrell}, {Fechner}, {Feiz}, {Feltzing}, {Ferreras},
  {Feuerstein}, {Feuillet}, {Finoguenov}, {Ford}, {Fotopoulou}, {Fouesneau},
  {Frenk}, {Frey}, {Gaessler}, {Geier}, {Gentile Fusillo}, {Gerhard},
  {Giannantonio}, {Giannone}, {Gibson}, {Gillingham},
  {Gonz{\'a}lez-Fern{\'a}ndez}, {Gonzalez-Solares}, {Gottloeber}, {Gould},
  {Grebel}, {Gueguen}, {Guiglion}, {Haehnelt}, {Hahn}, {Hansen}, {Hartman},
  {Hauptner}, {Hawkins}, {Haynes}, {Haynes}, {Heiter}, {Helmi}, {Aguayo},
  {Hewett}, {Hinton}, {Hobbs}, {Hoenig}, {Hofman}, {Hook}, {Hopgood},
  {Hopkins}, {Hourihane}, {Howes}, {Howlett}, {Huet}, {Irwin}, {Iwert},
  {Jablonka}, {Jahn}, {Jahnke}, {Jarno}, {Jin}, {Jofre}, {Johl}, {Jones},
  {J{\"o}nsson}, {Jordan}, {Karovicova}, {Khalatyan}, {Kelz}, {Kennicutt},
  {King}, {Kitaura}, {Klar}, {Klauser}, {Kneib}, {Koch}, {Koposov},
  {Kordopatis}, {Korn}, {Kosmalski}, {Kotak}, {Kovalev}, {Kreckel}, {Kripak},
  {Krumpe}, {Kuijken}, {Kunder}, {Kushniruk}, {Lam}, {Lamer}, {Laurent},
  {Lawrence}, {Lehmitz}, {Lemasle}, {Lewis}, {Li}, {Lidman}, {Lind}, {Liske},
  {Lizon}, {Loveday}, {Ludwig}, {McDermid}, {Maguire}, {Mainieri}, {Mali}, \&
  {Mandel}}]{4MOST_2019}
{de Jong}, R.~S., {Agertz}, O., {Berbel}, A.~A., {et~al.} 2019, The Messenger,
  175, 3

\bibitem[{{Dewi} {et~al.}(2006){Dewi}, {Podsiadlowski}, \& {Sena}}]{Dewi2006}
{Dewi}, J.~D.~M., {Podsiadlowski}, P., \& {Sena}, A. 2006, \mnras, 368, 1742

\bibitem[{{Dorsch} {et~al.}(2024){Dorsch}, {Jeffery}, {Philip Monai}, {Tout},
  {Snowdon}, {Monageng}, {Scott}, {Miszalski}, \& {Woolf}}]{Dorsch2024}
{Dorsch}, M., {Jeffery}, C.~S., {Philip Monai}, A., {et~al.} 2024, \aap, 691,
  A165

\bibitem[{{Dorsch} {et~al.}(2022){Dorsch}, {Reindl}, {Pelisoli}, {Heber},
  {Geier}, {Istrate}, \& {Justham}}]{Matti2022}
{Dorsch}, M., {Reindl}, N., {Pelisoli}, I., {et~al.} 2022, \aap, 658, L9

\bibitem[{{Duch{\^e}ne} \& {Kraus}(2013)}]{Duchene2013}
{Duch{\^e}ne}, G. \& {Kraus}, A. 2013, \araa, 51, 269

\bibitem[{{Eggleton}(1983)}]{EggletonRL}
{Eggleton}, P.~P. 1983, \apj, 268, 368

\bibitem[{Eggleton \& Yakut(2017)}]{60containinggiants}
Eggleton, P.~P. \& Yakut, K. 2017, Monthly Notices of the Royal Astronomical
  Society, 468, 3533

\bibitem[{{El-Badry} {et~al.}(2021){El-Badry}, {Rix}, \&
  {Heintz}}]{El-Badry2021}
{El-Badry}, K., {Rix}, H.-W., \& {Heintz}, T.~M. 2021, \mnras, 506, 2269

\bibitem[{{Ferrario} {et~al.}(2009){Ferrario}, {Pringle}, {Tout}, \&
  {Wickramasinghe}}]{Ferrario2009}
{Ferrario}, L., {Pringle}, J.~E., {Tout}, C.~A., \& {Wickramasinghe}, D.~T.
  2009, \mnras, 400, L71

\bibitem[{{Fitzpatrick} {et~al.}(2019){Fitzpatrick}, {Massa}, {Gordon},
  {Bohlin}, \& {Clayton}}]{Fitzpatrick2019}
{Fitzpatrick}, E.~L., {Massa}, D., {Gordon}, K.~D., {Bohlin}, R., \& {Clayton},
  G.~C. 2019, \apj, 886, 108

\bibitem[{Foreman-Mackey {et~al.}(2013)Foreman-Mackey, Hogg, Lang, \&
  Goodman}]{mcmc2013}
Foreman-Mackey, D., Hogg, D.~W., Lang, D., \& Goodman, J. 2013, Publications of
  the Astronomical Society of the Pacific, 125, 306–312

\bibitem[{{Gaia Collaboration} {et~al.}(2023{\natexlab{a}}){Gaia
  Collaboration}, {Arenou}, {Babusiaux}, {Barstow}, {Faigler}, {Jorissen},
  {Kervella}, {Mazeh}, {Mowlavi}, {Panuzzo}, {Sahlmann}, {Shahaf}, {Sozzetti},
  {Bauchet}, {Damerdji}, {Gavras}, {Giacobbe}, {Gosset}, {Halbwachs}, {Holl},
  {Lattanzi}, {Leclerc}, {Morel}, {Pourbaix}, {Re Fiorentin}, {Sadowski},
  {S{\'e}gransan}, {Siopis}, {Teyssier}, {Zwitter}, {Planquart}, {Brown},
  {Vallenari}, {Prusti}, {de Bruijne}, {Biermann}, {Creevey}, {Ducourant},
  {Evans}, {Eyer}, {Guerra}, {Hutton}, {Jordi}, {Klioner}, {Lammers},
  {Lindegren}, {Luri}, {Mignard}, {Panem}, {Randich}, {Sartoretti}, {Soubiran},
  {Tanga}, {Walton}, {Bailer-Jones}, {Bastian}, {Drimmel}, {Jansen}, {Katz},
  {van Leeuwen}, {Bakker}, {Cacciari}, {Casta{\~n}eda}, {De Angeli},
  {Fabricius}, {Fouesneau}, {Fr{\'e}mat}, {Galluccio}, {Guerrier}, {Heiter},
  {Masana}, {Messineo}, {Nicolas}, {Nienartowicz}, {Pailler}, {Riclet}, {Roux},
  {Seabroke}, {Sordo}, {Th{\'e}venin}, {Gracia-Abril}, {Portell}, {Altmann},
  {Andrae}, {Audard}, {Bellas-Velidis}, {Benson}, {Berthier}, {Blomme},
  {Burgess}, {Busonero}, {Busso}, {C{\'a}novas}, {Carry}, {Cellino}, {Cheek},
  {Clementini}, {Davidson}, {de Teodoro}, {Nu{\~n}ez Campos}, {Delchambre},
  {Dell'Oro}, {Esquej}, {Fern{\'a}ndez-Hern{\'a}ndez}, {Fraile}, {Garabato},
  {Garc{\'\i}a-Lario}, {Haigron}, {Hambly}, {Harrison}, {Hern{\'a}ndez},
  {Hestroffer}, {Hodgkin}, {Jan{\ss}en}, {Jevardat de Fombelle}, {Jordan},
  {Krone-Martins}, {Lanzafame}, {L{\"o}ffler}, {Marchal}, {Marrese},
  {Moitinho}, {Muinonen}, {Osborne}, {Pancino}, {Pauwels}, {Recio-Blanco},
  {Reyl{\'e}}, {Riello}, {Rimoldini}, {Roegiers}, {Rybizki}, {Sarro}, {Smith},
  {Utrilla}, {van Leeuwen}, {Abbas}, {{\'A}brah{\'a}m}, {Abreu Aramburu},
  {Aerts}, {Aguado}, {Ajaj}, {Aldea-Montero}, {Altavilla}, {{\'A}lvarez},
  {Alves}, {Anders}, {Anderson}, {Anglada Varela}, {Antoja}, {Baines}, {Baker},
  {Balaguer-N{\'u}{\~n}ez}, {Balbinot}, {Balog}, {Barache}, {Barbato},
  {Barros}, {Bartolom{\'e}}, {Bassilana}, {Becciani}, {Bellazzini},
  {Berihuete}, {Bernet}, {Bertone}, {Bianchi}, {Binnenfeld}, {Blanco-Cuaresma},
  {Blazere}, {Boch}, {Bombrun}, {Bossini}, {Bouquillon}, {Bragaglia},
  {Bramante}, {Breedt}, {Bressan}, {Brouillet}, {Brugaletta}, {Bucciarelli},
  {Burlacu}, {Butkevich}, {Buzzi}, {Caffau}, {Cancelliere}, {Cantat-Gaudin},
  {Carballo}, {Carlucci}, {Carnerero}, {Carrasco}, {Casamiquela}, {Castellani},
  {Castro-Ginard}, {Chaoul}, {Charlot}, {Chemin}, {Chiaramida}, {Chiavassa},
  {Chornay}, \& {Comoretto}}]{GaiaDR3_binary}
{Gaia Collaboration}, {Arenou}, F., {Babusiaux}, C., {et~al.}
  2023{\natexlab{a}}, \aap, 674, A34

\bibitem[{{Gaia Collaboration} {et~al.}(2018){Gaia Collaboration}, {Brown},
  {Vallenari}, {Prusti}, {de Bruijne}, {Babusiaux}, {Bailer-Jones}, {Biermann},
  {Evans}, {Eyer}, {Jansen}, {Jordi}, {Klioner}, {Lammers}, {Lindegren},
  {Luri}, {Mignard}, {Panem}, {Pourbaix}, {Randich}, {Sartoretti}, {Siddiqui},
  {Soubiran}, {van Leeuwen}, {Walton}, {Arenou}, {Bastian}, {Cropper},
  {Drimmel}, {Katz}, {Lattanzi}, {Bakker}, {Cacciari}, {Casta{\~n}eda},
  {Chaoul}, {Cheek}, {De Angeli}, {Fabricius}, {Guerra}, {Holl}, {Masana},
  {Messineo}, {Mowlavi}, {Nienartowicz}, {Panuzzo}, {Portell}, {Riello},
  {Seabroke}, {Tanga}, {Th{\'e}venin}, {Gracia-Abril}, {Comoretto},
  {Garcia-Reinaldos}, {Teyssier}, {Altmann}, {Andrae}, {Audard},
  {Bellas-Velidis}, {Benson}, {Berthier}, {Blomme}, {Burgess}, {Busso},
  {Carry}, {Cellino}, {Clementini}, {Clotet}, {Creevey}, {Davidson}, {De
  Ridder}, {Delchambre}, {Dell'Oro}, {Ducourant},
  {Fern{\'a}ndez-Hern{\'a}ndez}, {Fouesneau}, {Fr{\'e}mat}, {Galluccio},
  {Garc{\'\i}a-Torres}, {Gonz{\'a}lez-N{\'u}{\~n}ez}, {Gonz{\'a}lez-Vidal},
  {Gosset}, {Guy}, {Halbwachs}, {Hambly}, {Harrison}, {Hern{\'a}ndez},
  {Hestroffer}, {Hodgkin}, {Hutton}, {Jasniewicz}, {Jean-Antoine-Piccolo},
  {Jordan}, {Korn}, {Krone-Martins}, {Lanzafame}, {Lebzelter}, {L{\"o}ffler},
  {Manteiga}, {Marrese}, {Mart{\'\i}n-Fleitas}, {Moitinho}, {Mora}, {Muinonen},
  {Osinde}, {Pancino}, {Pauwels}, {Petit}, {Recio-Blanco}, {Richards},
  {Rimoldini}, {Robin}, {Sarro}, {Siopis}, {Smith}, {Sozzetti}, {S{\"u}veges},
  {Torra}, {van Reeven}, {Abbas}, {Abreu Aramburu}, {Accart}, {Aerts},
  {Altavilla}, {{\'A}lvarez}, {Alvarez}, {Alves}, {Anderson}, {Andrei},
  {Anglada Varela}, {Antiche}, {Antoja}, {Arcay}, {Astraatmadja}, {Bach},
  {Baker}, {Balaguer-N{\'u}{\~n}ez}, {Balm}, {Barache}, {Barata}, {Barbato},
  {Barblan}, {Barklem}, {Barrado}, {Barros}, {Barstow}, {Bartholom{\'e}
  Mu{\~n}oz}, {Bassilana}, {Becciani}, {Bellazzini}, {Berihuete}, {Bertone},
  {Bianchi}, {Bienaym{\'e}}, {Blanco-Cuaresma}, {Boch}, {Boeche}, {Bombrun},
  {Borrachero}, {Bossini}, {Bouquillon}, {Bourda}, {Bragaglia}, {Bramante},
  {Breddels}, {Bressan}, {Brouillet}, {Br{\"u}semeister}, {Brugaletta},
  {Bucciarelli}, {Burlacu}, {Busonero}, {Butkevich}, {Buzzi}, {Caffau},
  {Cancelliere}, {Cannizzaro}, {Cantat-Gaudin}, {Carballo}, {Carlucci},
  {Carrasco}, {Casamiquela}, {Castellani}, {Castro-Ginard}, {Charlot},
  {Chemin}, {Chiavassa}, {Cocozza}, {Costigan}, {Cowell}, {Crifo}, {Crosta},
  {Crowley}, {Cuypers}, {Dafonte}, {Damerdji}, {Dapergolas}, {David}, {David},
  {de Laverny}, \& {De Luise}}]{GaiaDR2}
{Gaia Collaboration}, {Brown}, A.~G.~A., {Vallenari}, A., {et~al.} 2018, \aap,
  616, A1

\bibitem[{{Gaia Collaboration} {et~al.}(2023{\natexlab{b}}){Gaia
  Collaboration}, {Vallenari}, {Brown}, {Prusti}, {de Bruijne}, {Arenou},
  {Babusiaux}, {Biermann}, {Creevey}, {Ducourant}, {Evans}, {Eyer}, {Guerra},
  {Hutton}, {Jordi}, {Klioner}, {Lammers}, {Lindegren}, {Luri}, {Mignard},
  {Panem}, {Pourbaix}, {Randich}, {Sartoretti}, {Soubiran}, {Tanga}, {Walton},
  {Bailer-Jones}, {Bastian}, {Drimmel}, {Jansen}, {Katz}, {Lattanzi}, {van
  Leeuwen}, {Bakker}, {Cacciari}, {Casta{\~n}eda}, {De Angeli}, {Fabricius},
  {Fouesneau}, {Fr{\'e}mat}, {Galluccio}, {Guerrier}, {Heiter}, {Masana},
  {Messineo}, {Mowlavi}, {Nicolas}, {Nienartowicz}, {Pailler}, {Panuzzo},
  {Riclet}, {Roux}, {Seabroke}, {Sordo}, {Th{\'e}venin}, {Gracia-Abril},
  {Portell}, {Teyssier}, {Altmann}, {Andrae}, {Audard}, {Bellas-Velidis},
  {Benson}, {Berthier}, {Blomme}, {Burgess}, {Busonero}, {Busso},
  {C{\'a}novas}, {Carry}, {Cellino}, {Cheek}, {Clementini}, {Damerdji},
  {Davidson}, {de Teodoro}, {Nu{\~n}ez Campos}, {Delchambre}, {Dell'Oro},
  {Esquej}, {Fern{\'a}ndez-Hern{\'a}ndez}, {Fraile}, {Garabato},
  {Garc{\'\i}a-Lario}, {Gosset}, {Haigron}, {Halbwachs}, {Hambly}, {Harrison},
  {Hern{\'a}ndez}, {Hestroffer}, {Hodgkin}, {Holl}, {Jan{\ss}en}, {Jevardat de
  Fombelle}, {Jordan}, {Krone-Martins}, {Lanzafame}, {L{\"o}ffler}, {Marchal},
  {Marrese}, {Moitinho}, {Muinonen}, {Osborne}, {Pancino}, {Pauwels},
  {Recio-Blanco}, {Reyl{\'e}}, {Riello}, {Rimoldini}, {Roegiers}, {Rybizki},
  {Sarro}, {Siopis}, {Smith}, {Sozzetti}, {Utrilla}, {van Leeuwen}, {Abbas},
  {{\'A}brah{\'a}m}, {Abreu Aramburu}, {Aerts}, {Aguado}, {Ajaj},
  {Aldea-Montero}, {Altavilla}, {{\'A}lvarez}, {Alves}, {Anders}, {Anderson},
  {Anglada Varela}, {Antoja}, {Baines}, {Baker}, {Balaguer-N{\'u}{\~n}ez},
  {Balbinot}, {Balog}, {Barache}, {Barbato}, {Barros}, {Barstow},
  {Bartolom{\'e}}, {Bassilana}, {Bauchet}, {Becciani}, {Bellazzini},
  {Berihuete}, {Bernet}, {Bertone}, {Bianchi}, {Binnenfeld}, {Blanco-Cuaresma},
  {Blazere}, {Boch}, {Bombrun}, {Bossini}, {Bouquillon}, {Bragaglia},
  {Bramante}, {Breedt}, {Bressan}, {Brouillet}, {Brugaletta}, {Bucciarelli},
  {Burlacu}, {Butkevich}, {Buzzi}, {Caffau}, {Cancelliere}, {Cantat-Gaudin},
  {Carballo}, {Carlucci}, {Carnerero}, {Carrasco}, {Casamiquela}, {Castellani},
  {Castro-Ginard}, {Chaoul}, {Charlot}, {Chemin}, {Chiaramida}, {Chiavassa},
  {Chornay}, {Comoretto}, {Contursi}, {Cooper}, {Cornez}, {Cowell}, {Crifo},
  {Cropper}, {Crosta}, {Crowley}, {Dafonte}, {Dapergolas}, {David}, {David},
  {de Laverny}, {De Luise}, \& {De March}}]{GaiaDR3}
{Gaia Collaboration}, {Vallenari}, A., {Brown}, A.~G.~A., {et~al.}
  2023{\natexlab{b}}, \aap, 674, A1

\bibitem[{{Green} {et~al.}(2019){Green}, {Schlafly}, {Zucker}, {Speagle}, \&
  {Finkbeiner}}]{Green2019}
{Green}, G.~M., {Schlafly}, E., {Zucker}, C., {Speagle}, J.~S., \&
  {Finkbeiner}, D. 2019, \apj, 887, 93

\bibitem[{{Han}(1998)}]{Han1998}
{Han}, Z. 1998, \mnras, 296, 1019

\bibitem[{{Han} {et~al.}(2003){Han}, {Podsiadlowski}, {Maxted}, \&
  {Marsh}}]{Han2003}
{Han}, Z., {Podsiadlowski}, P., {Maxted}, P.~F.~L., \& {Marsh}, T.~R. 2003,
  \mnras, 341, 669

\bibitem[{{Han} {et~al.}(2002){Han}, {Podsiadlowski}, {Maxted}, {Marsh}, \&
  {Ivanova}}]{Han2002}
{Han}, Z., {Podsiadlowski}, P., {Maxted}, P.~F.~L., {Marsh}, T.~R., \&
  {Ivanova}, N. 2002, \mnras, 336, 449

\bibitem[{{Heber}(2024)}]{Heber2024}
{Heber}, U. 2024, arXiv e-prints, arXiv:2410.11663

\bibitem[{{Heber} {et~al.}(2018){Heber}, {Irrgang}, \&
  {Schaffenroth}}]{hebersed}
{Heber}, U., {Irrgang}, A., \& {Schaffenroth}, J. 2018, Open Astronomy, 27, 35

\bibitem[{{Henden} {et~al.}(2016){Henden}, {Templeton}, {Terrell}, {Smith},
  {Levine}, \& {Welch}}]{Henden_apass9}
{Henden}, A.~A., {Templeton}, M., {Terrell}, D., {et~al.} 2016, {VizieR Online
  Data Catalog: AAVSO Photometric All Sky Survey (APASS) DR9 (Henden+, 2016)},
  VizieR On-line Data Catalog: II/336. Originally published in:
  2015AAS...22533616H

\bibitem[{{Houck} \& {Denicola}(2000)}]{ISIS}
{Houck}, J.~C. \& {Denicola}, L.~A. 2000, in Astronomical Society of the
  Pacific Conference Series, Vol. 216, Astronomical Data Analysis Software and
  Systems IX, ed. N.~{Manset}, C.~{Veillet}, \& D.~{Crabtree}, 591

\bibitem[{{Iben} \& {Tutukov}(1984)}]{Iben1984}
{Iben}, Jr., I. \& {Tutukov}, A.~V. 1984, \apj, 284, 719

\bibitem[{{Ivanova} {et~al.}(2013){Ivanova}, {Justham}, {Chen}, {De Marco},
  {Fryer}, {Gaburov}, {Ge}, {Glebbeek}, {Han}, {Li}, {Lu}, {Marsh},
  {Podsiadlowski}, {Potter}, {Soker}, {Taam}, {Tauris}, {van den Heuvel}, \&
  {Webbink}}]{Ivanova2013}
{Ivanova}, N., {Justham}, S., {Chen}, X., {et~al.} 2013, \aapr, 21, 59

\bibitem[{{Justham} {et~al.}(2011){Justham}, {Podsiadlowski}, \&
  {Han}}]{Justham_HerichsdO}
{Justham}, S., {Podsiadlowski}, P., \& {Han}, Z. 2011, \mnras, 410, 984

\bibitem[{{Kab{\'a}th} {et~al.}(2020){Kab{\'a}th}, {Skarka}, {Sabotta},
  {Guenther}, {Jones}, {Klocov{\'a}}, {{\v{S}}ubjak}, {{\v{Z}}{\'a}k},
  {{\v{S}}pokov{\'a}}, {Bla{\v{z}}ek}, {Dvo{\v{r}}{\'a}kov{\'a}}, {Dupkala},
  {Fuchs}, {Hatzes}, {Kortusov{\'a}}, {Novotn{\'y}}, {Pl{\'a}valov{\'a}},
  {{\v{R}}ezba}, {Sloup}, {{\v{S}}koda}, \& {{\v{S}}lechta}}]{Kabath2020}
{Kab{\'a}th}, P., {Skarka}, M., {Sabotta}, S., {et~al.} 2020, \pasp, 132,
  035002

\bibitem[{{Koubsk{\'y}} {et~al.}(2004){Koubsk{\'y}}, {Mayer}, {{\v{C}}{\'a}p},
  {{\v{Z}}{\v{d}}{\'a}rsk{\'y}}, {Zeman}, {P{\'\i}na}, \& {Melich}}]{OES}
{Koubsk{\'y}}, P., {Mayer}, P., {{\v{C}}{\'a}p}, J., {et~al.} 2004,
  Publications of the Astronomical Institute of the Czechoslovak Academy of
  Sciences, 92, 37

\bibitem[{{Kounkel} {et~al.}(2021){Kounkel}, {Covey}, {Stassun},
  {Price-Whelan}, {Holtzman}, {Chojnowski}, {Longa-Pe{\~n}a},
  {Rom{\'a}n-Z{\'u}{\~n}iga}, {Hernandez}, {Serna}, {Badenes}, {De Lee},
  {Majewski}, {Stringfellow}, {Kratter}, {Moe}, {Frinchaboy}, {Beaton},
  {Fern{\'a}ndez-Trincado}, {Mahadevan}, {Minniti}, {Beers}, {Schneider},
  {Barba}, {Brownstein}, {Garc{\'\i}a-Hern{\'a}ndez}, {Pan}, \&
  {Bizyaev}}]{APOGEE}
{Kounkel}, M., {Covey}, K.~R., {Stassun}, K.~G., {et~al.} 2021, \aj, 162, 184

\bibitem[{Kurucz(2013)}]{Atlas}
Kurucz, R. 2013, Astrophysics Source Code Library, 03024

\bibitem[{{Kurucz}(1993)}]{SYNTHE}
{Kurucz}, R.~L. 1993, {SYNTHE spectrum synthesis programs and line data}

\bibitem[{{Li} {et~al.}(2019){Li}, {Chen}, {Chen}, \& {Han}}]{Li2019}
{Li}, Z., {Chen}, X., {Chen}, H.-L., \& {Han}, Z. 2019, \apj, 871, 148

\bibitem[{{Lindegren} {et~al.}(2021){Lindegren}, {Bastian}, {Biermann},
  {Bombrun}, {de Torres}, {Gerlach}, {Geyer}, {Hern{\'a}ndez}, {Hilger},
  {Hobbs}, {Klioner}, {Lammers}, {McMillan}, {Ramos-Lerate},
  {Steidelm{\"u}ller}, {Stephenson}, \& {van Leeuwen}}]{Lindegren2021}
{Lindegren}, L., {Bastian}, U., {Biermann}, M., {et~al.} 2021, \aap, 649, A4

\bibitem[{{Lisker} {et~al.}(2004){Lisker}, {Heber}, {Napiwotzki}, {Christlieb},
  {Reimers}, \& {Homeier}}]{HE03}
{Lisker}, T., {Heber}, U., {Napiwotzki}, R., {et~al.} 2004, \apss, 291, 351

\bibitem[{{Munday} {et~al.}(2024){Munday}, {Pelisoli}, {Tremblay}, {Marsh},
  {Nelemans}, {B{\'e}dard}, {Toonen}, {Breedt}, {Cunningham}, {O'Brien}, \&
  {Dawson}}]{Munday2024}
{Munday}, J., {Pelisoli}, I., {Tremblay}, P.~E., {et~al.} 2024, \mnras, 532,
  2534

\bibitem[{{Nelemans} \& {Tout}(2005)}]{Nelemans2005}
{Nelemans}, G. \& {Tout}, C.~A. 2005, \mnras, 356, 753

\bibitem[{{Offner} {et~al.}(2023){Offner}, {Moe}, {Kratter}, {Sadavoy},
  {Jensen}, \& {Tobin}}]{Offner2023}
{Offner}, S.~S.~R., {Moe}, M., {Kratter}, K.~M., {et~al.} 2023, in Astronomical
  Society of the Pacific Conference Series, Vol. 534, Protostars and Planets
  VII, ed. S.~{Inutsuka}, Y.~{Aikawa}, T.~{Muto}, K.~{Tomida}, \& M.~{Tamura},
  275

\bibitem[{{Paczynski}(1976)}]{Paczynski1976}
{Paczynski}, B. 1976, in IAU Symposium, Vol.~73, Structure and Evolution of
  Close Binary Systems, ed. P.~{Eggleton}, S.~{Mitton}, \& J.~{Whelan}, 75

\bibitem[{{Pakmor} {et~al.}(2024){Pakmor}, {Pelisoli}, {Justham},
  {Rajamuthukumar}, {R{\"o}pke}, {Schneider}, {de Mink}, {Ohlmann},
  {Podsiadlowski}, {Mor{\'a}n-Fraile}, {Vetter}, \& {Andrassy}}]{Pakmor2024}
{Pakmor}, R., {Pelisoli}, I., {Justham}, S., {et~al.} 2024, \aap, 691, A179

\bibitem[{{Pelisoli} {et~al.}(2022){Pelisoli}, {Dorsch}, {Heber},
  {G{\"a}nsicke}, {Geier}, {Kupfer}, {N{\'e}meth}, {Scaringi}, \&
  {Schaffenroth}}]{Pelisoli2022}
{Pelisoli}, I., {Dorsch}, M., {Heber}, U., {et~al.} 2022, \mnras, 515, 2496

\bibitem[{{Pietrinferni} {et~al.}(2021){Pietrinferni}, {Hidalgo}, {Cassisi},
  {Salaris}, {Savino}, {Mucciarelli}, {Verma}, {Silva Aguirre}, {Aparicio}, \&
  {Ferguson}}]{BaSTI2021}
{Pietrinferni}, A., {Hidalgo}, S., {Cassisi}, S., {et~al.} 2021, \apj, 908, 102

\bibitem[{{Pribulla} {et~al.}(2015){Pribulla}, {Garai}, {Hamb{\'a}lek},
  {Koll{\'a}r}, {Kom{\v{z}}{\'\i}k}, {Kundra}, {Nedoro{\v{s}}{\v{c}}{\'\i}k},
  {Seker{\'a}{\v{s}}}, \& {Va{\v{n}}ko}}]{Pribulla2015}
{Pribulla}, T., {Garai}, Z., {Hamb{\'a}lek}, L., {et~al.} 2015, Astronomische
  Nachrichten, 336, 682

\bibitem[{{Raskin}(2011)}]{HERMES}
{Raskin}, G. 2011, PhD thesis, Katholieke University of Leuven, Astronomical
  Institute

\bibitem[{{Rawls} {et~al.}(2016){Rawls}, {Gaulme}, {McKeever}, {Jackiewicz},
  {Orosz}, {Corsaro}, {Beck}, {Mosser}, {Latham}, \& {Latham}}]{doubleRGosc}
{Rawls}, M.~L., {Gaulme}, P., {McKeever}, J., {et~al.} 2016, \apj, 818, 108

\bibitem[{{Ricker} {et~al.}(2015){Ricker}, {Winn}, {Vanderspek}, {Latham},
  {Bakos}, {Bean}, {Berta-Thompson}, {Brown}, {Buchhave}, {Butler}, {Butler},
  {Chaplin}, {Charbonneau}, {Christensen-Dalsgaard}, {Clampin}, {Deming},
  {Doty}, {De Lee}, {Dressing}, {Dunham}, {Endl}, {Fressin}, {Ge}, {Henning},
  {Holman}, {Howard}, {Ida}, {Jenkins}, {Jernigan}, {Johnson}, {Kaltenegger},
  {Kawai}, {Kjeldsen}, {Laughlin}, {Levine}, {Lin}, {Lissauer}, {MacQueen},
  {Marcy}, {McCullough}, {Morton}, {Narita}, {Paegert}, {Palle}, {Pepe},
  {Pepper}, {Quirrenbach}, {Rinehart}, {Sasselov}, {Sato}, {Seager},
  {Sozzetti}, {Stassun}, {Sullivan}, {Szentgyorgyi}, {Torres}, {Udry}, \&
  {Villasenor}}]{TESS_2015}
{Ricker}, G.~R., {Winn}, J.~N., {Vanderspek}, R., {et~al.} 2015, Journal of
  Astronomical Telescopes, Instruments, and Systems, 1, 014003

\bibitem[{{Riello} {et~al.}(2021){Riello}, {De Angeli}, {Evans}, {Montegriffo},
  {Carrasco}, {Busso}, {Palaversa}, {Burgess}, {Diener}, {Davidson}, {Rowell},
  {Fabricius}, {Jordi}, {Bellazzini}, {Pancino}, {Harrison}, {Cacciari}, {van
  Leeuwen}, {Hambly}, {Hodgkin}, {Osborne}, {Altavilla}, {Barstow}, {Brown},
  {Castellani}, {Cowell}, {De Luise}, {Gilmore}, {Giuffrida}, {Hidalgo},
  {Holland}, {Marinoni}, {Pagani}, {Piersimoni}, {Pulone}, {Ragaini}, {Rainer},
  {Richards}, {Sanna}, {Walton}, {Weiler}, \& {Yoldas}}]{Riello2021}
{Riello}, M., {De Angeli}, F., {Evans}, D.~W., {et~al.} 2021, \aap, 649, A3

\bibitem[{{Ruiz-Dern} {et~al.}(2018){Ruiz-Dern}, {Babusiaux}, {Arenou},
  {Turon}, \& {Lallement}}]{Ruiz-Dern2018}
{Ruiz-Dern}, L., {Babusiaux}, C., {Arenou}, F., {Turon}, C., \& {Lallement}, R.
  2018, \aap, 609, A116

\bibitem[{{Schlafly} \& {Finkbeiner}(2011)}]{SF2011}
{Schlafly}, E.~F. \& {Finkbeiner}, D.~P. 2011, \apj, 737, 103

\bibitem[{{Schlegel} {et~al.}(1998){Schlegel}, {Finkbeiner}, \&
  {Davis}}]{SFD1998}
{Schlegel}, D.~J., {Finkbeiner}, D.~P., \& {Davis}, M. 1998, \apj, 500, 525

\bibitem[{{Schwab}(2018)}]{Schwab2018}
{Schwab}, J. 2018, \mnras, 476, 5303

\bibitem[{Shao \& Li(2014)}]{Be}
Shao, Y. \& Li, X.-D. 2014, The Astrophysical Journal, 796, 37

\bibitem[{{Skrutskie} {et~al.}(2006){Skrutskie}, {Cutri}, {Stiening},
  {Weinberg}, {Schneider}, {Carpenter}, {Beichman}, {Capps}, {Chester},
  {Elias}, {Huchra}, {Liebert}, {Lonsdale}, {Monet}, {Price}, {Seitzer},
  {Jarrett}, {Kirkpatrick}, {Gizis}, {Howard}, {Evans}, {Fowler}, {Fullmer},
  {Hurt}, {Light}, {Kopan}, {Marsh}, {McCallon}, {Tam}, {Van Dyk}, \&
  {Wheelock}}]{Skrutskie2006_2MASS}
{Skrutskie}, M.~F., {Cutri}, R.~M., {Stiening}, R., {et~al.} 2006, \aj, 131,
  1163

\bibitem[{{Taam} \& {Ricker}(2010)}]{Taam2010}
{Taam}, R.~E. \& {Ricker}, P.~M. 2010, \nar, 54, 65

\bibitem[{{Tody}(1986)}]{iraf}
{Tody}, D. 1986, in Society of Photo-Optical Instrumentation Engineers (SPIE)
  Conference Series, Vol. 627, Instrumentation in astronomy VI, ed. D.~L.
  {Crawford}, 733

\bibitem[{Tody(1993)}]{Tody1993}
Tody, D. 1993, 26, 791

\bibitem[{{Tout} \& {Eggleton}(1988)}]{Tout1988}
{Tout}, C.~A. \& {Eggleton}, P.~P. 1988, \mnras, 231, 823

\bibitem[{{Traven} {et~al.}(2020){Traven}, {Feltzing}, {Merle}, {Van der
  Swaelmen}, {{\v{C}}otar}, {Church}, {Zwitter}, {Ting}, {Sahlholdt},
  {Asplund}, {Bland-Hawthorn}, {De Silva}, {Freeman}, {Martell}, {Sharma},
  {Zucker}, {Buder}, {Casey}, {D'Orazi}, {Kos}, {Lewis}, {Lin}, {Lind},
  {Simpson}, {Stello}, {Munari}, \& {Wittenmyer}}]{GALAH}
{Traven}, G., {Feltzing}, S., {Merle}, T., {et~al.} 2020, \aap, 638, A145

\bibitem[{{Tutukov} \& {Fedorova}(2010)}]{Tutukov2010}
{Tutukov}, A.~V. \& {Fedorova}, A.~V. 2010, Astronomy Reports, 54, 156

\bibitem[{{Uzundag} {et~al.}(2022){Uzundag}, {Jones}, {Vu{\v{c}}kovi{\'c}},
  {Vos}, {Bobrick}, \& {Paladini}}]{Uzundag2022}
{Uzundag}, M., {Jones}, M.~I., {Vu{\v{c}}kovi{\'c}}, M., {et~al.} 2022, \aap,
  668, A89

\bibitem[{{van der Sluys} {et~al.}(2006){van der Sluys}, {Verbunt}, \&
  {Pols}}]{vanderSluys2006}
{van der Sluys}, M.~V., {Verbunt}, F., \& {Pols}, O.~R. 2006, \aap, 460, 209

\bibitem[{Virtanen {et~al.}(2020)Virtanen, Gommers, Oliphant, Haberland, Reddy,
  Cournapeau, Burovski, Peterson, Weckesser, Bright, {van der Walt}, Brett,
  Wilson, Millman, Mayorov, Nelson, Jones, Kern, Larson, Carey, Polat, Feng,
  Moore, {VanderPlas}, Laxalde, Perktold, Cimrman, Henriksen, Quintero, Harris,
  Archibald, Ribeiro, Pedregosa, {van Mulbregt}, \& {SciPy 1.0
  Contributors}}]{2020Scipy}
Virtanen, P., Gommers, R., Oliphant, T.~E., {et~al.} 2020, Nature Methods, 17,
  261

\bibitem[{{Webbink}(1984)}]{Webbink1984}
{Webbink}, R.~F. 1984, \apj, 277, 355

\bibitem[{{Yu} {et~al.}(2021){Yu}, {Zhang}, \& {L{\"u}}}]{Yu2021}
{Yu}, J., {Zhang}, X., \& {L{\"u}}, G. 2021, \mnras, 504, 2670

\bibitem[{{Zhang} \& {Jeffery}(2012)}]{Zhang2012}
{Zhang}, X. \& {Jeffery}, C.~S. 2012, \mnras, 419, 452

\end{thebibliography}

\begin{appendix}

\begin{figure*}[h!]
\section{Additional material}
\centering
\begin{subfigure}{0.5\textwidth}
    \centering
    \includegraphics[width=\linewidth]{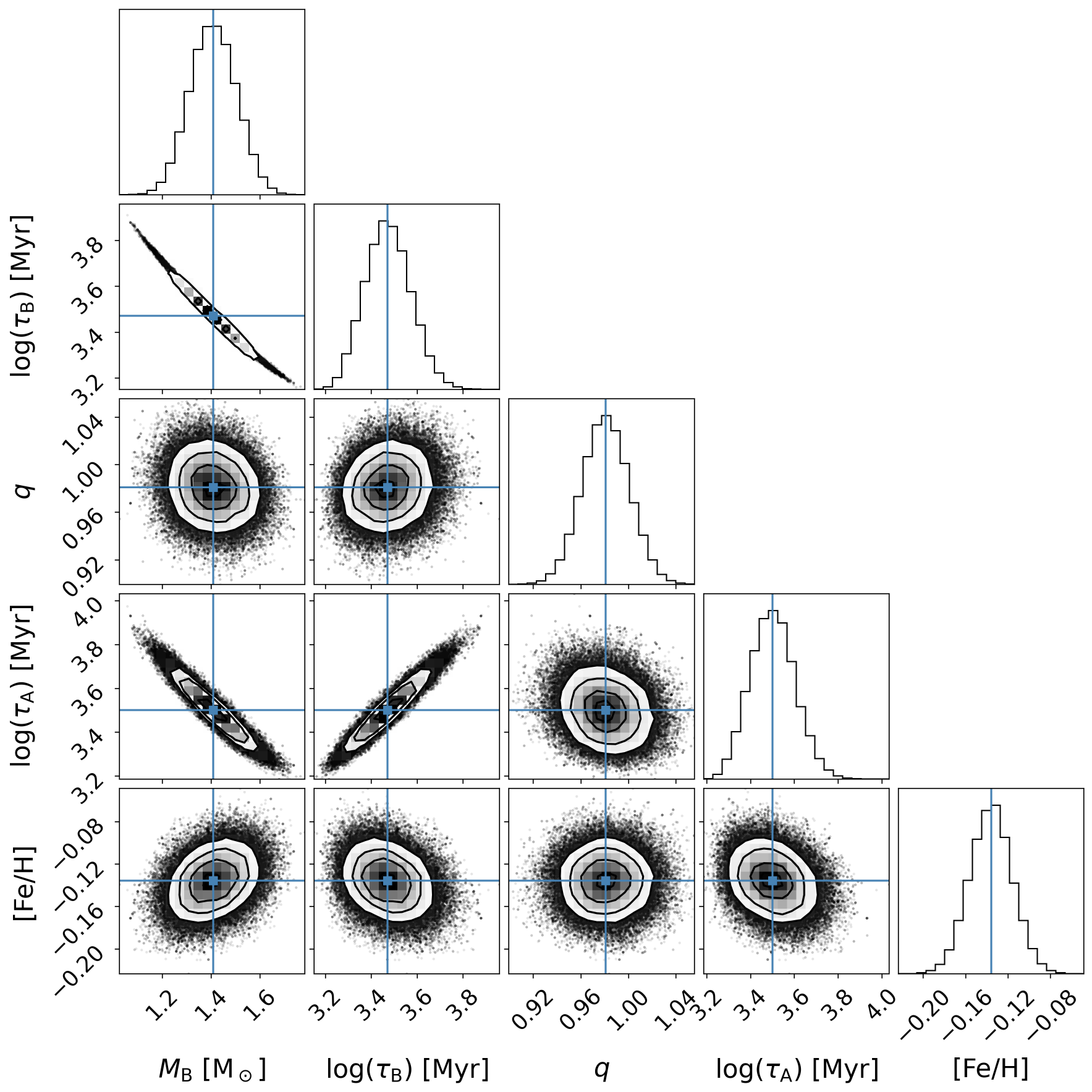}
    \label{fig:corner_fit}
\end{subfigure}%
\hfill
\begin{subfigure}{0.48\textwidth}
    \centering
    \includegraphics[width=\linewidth]{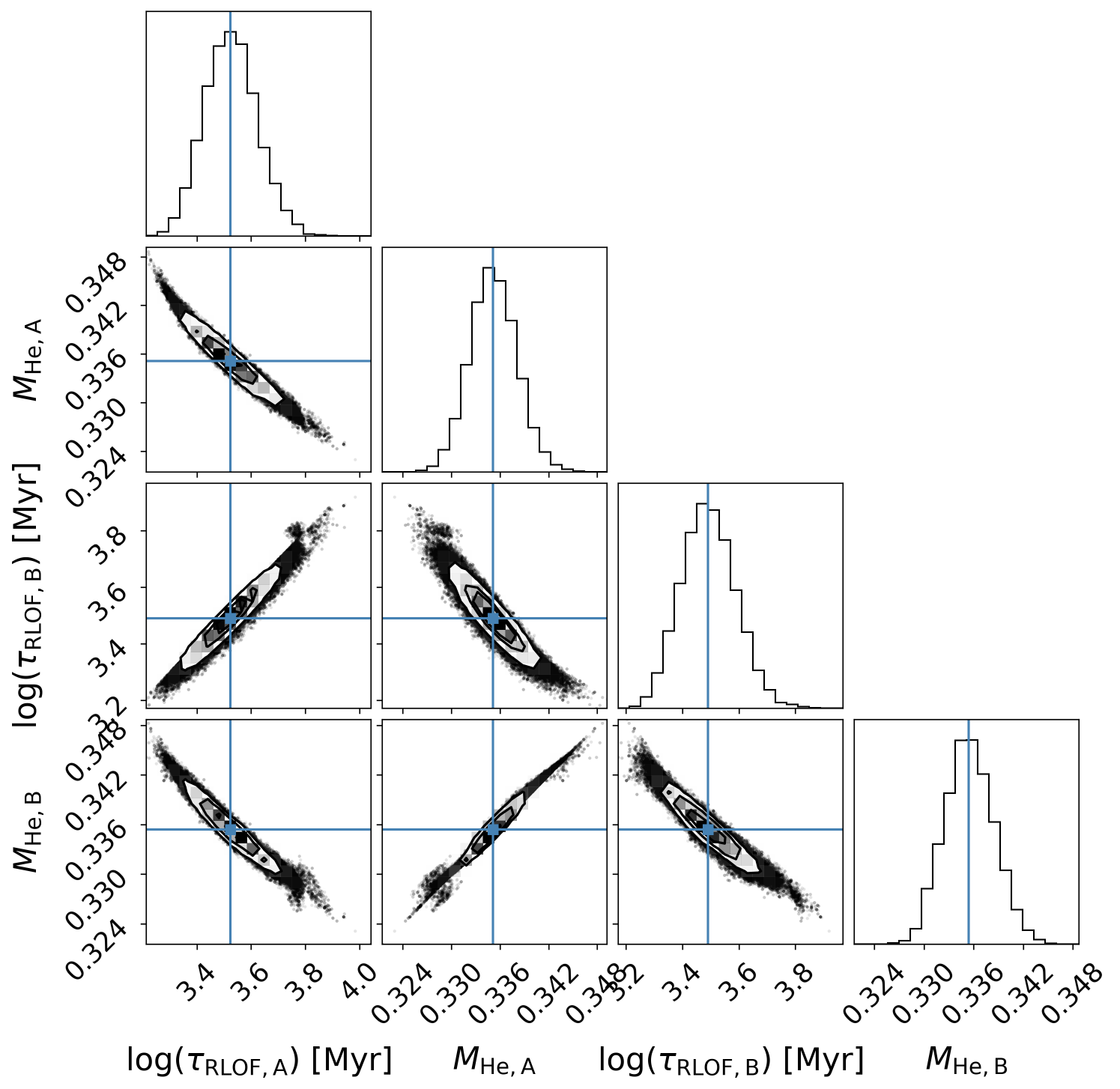}
    \label{fig:corner_RLOF}
\end{subfigure}
\caption{Corner plots displaying the posterior distributions of key parameters derived from the MCMC analysis. \textit{Left:} Parameter distributions from the best-fit evolutionary model. \textit{Right:} Posterior distributions at the point of RLOF.}
\label{fig:corner_plots}
\end{figure*}

\begin{table*}[h!]
    \centering
    \setstretch{1.1}
    \caption{\highlight{Heliocentrically corrected} radial velocity ($RV$) measurements.}
    \begin{tabular}{cccrcrc}
        \hline
        JD & Date& $t_{\mathrm{exp}}$ & $RV_\mathrm{A}$ & $RV_\mathrm{B}$ & Instrument \\
           &    (UTC)          & (hours)            & (\kms)          & (\kms)          &            \\
        \hline
        2459103.3356 & 2020-09-10 & 1.00 & 10.5  & -33.5 & OES \\
        2459448.5044 & 2021-08-22 & 0.37 & -11.0 & -11.0 & OES \\
        2459465.5555 & 2021-09-08 & 0.50 & -38.1 & 15.4  & OES \\
        2459743.5255 & 2022-06-13 & 0.29 & 12.5  & -34.4 & OES \\
        2459803.5541 & 2022-08-12 & 0.56 & -27.1 & 4.7   & OES \\
        2459804.4434 & 2022-08-12 & 0.56 & -24.9 & 2.4   & OES \\
        2459825.4714 & 2022-09-02 & 1.00 & 12.3  & -35.0 & OES \\
        2460180.4828 & 2023-08-23 & 1.00 & -14.7 & -1.0  & OES \\
        2460198.6307 & 2023-09-11 & 1.00 & -40.1 & 17.0  & OES \\
        2460209.5201 & 2023-09-22 & 1.00 & -25.2 & 2.8   & OES \\
        2460219.4013 & 2023-10-01 & 1.00 & 0.6   & -22.9 & OES \\
        2460553.4990 & 2024-08-30 & 0.50 & 11.9  & -34.6 & OES \\
        2460647.3722 & 2024-12-02 & 0.25 & 7.8   & -30.0 & HERMES \\
        2460650.2574 & 2024-12-05 & 0.78 & 5.1   & -27.2 & OES \\
        2460689.2031 & 2025-01-13 & 0.78 & -36.5 & 16.3  & OES \\
        2460717.2760 & 2025-02-10 & 0.78 & 12.5  & -34.0 & OES \\
        \hline
    \end{tabular}
    \tablefoot{Measurements are sorted by Julian date (JD). We estimate a uniform uncertainty of $\sigma_{\mathrm{RV}_\mathrm{A,B}}=1$ \kms\ for all radial velocity values, \highlight{which is consistent with the reduced $\chi^2$ of the RV curve fit}.}
    \label{tab:rv_measurements}
\end{table*}

\end{appendix}
\end{document}